\documentclass[11pt]{article}
\usepackage{amssymb,graphicx}
\newcommand\toline[1]{--#1}
\newcommand\php{^{\phantom{i}}}
\newcommand{\fract}[2]{{\textstyle\frac{#1}{#2}}}

\newcommand{\la}{z}

\newcommand\Z{{\mathbb Z}}
\newcommand\R{{\mathbb R}}
\newcommand\CH{{\cal H}}

\newcommand{\CP}{{\cal P}}
\newcommand{\CT}{{\cal T}}
\newcommand{\PT}{{\cal PT}}
\newcommand\eq{\begin{equation}}
\newcommand\en{\end{equation}}
\newcommand\bea{\begin{eqnarray}}
\newcommand\eea{\end{eqnarray}}
\newcommand\nn{\nonumber}
\newcommand\ba{\(\begin{array}}
\newcommand\ea{\end{array}\)}

\newcommand{\resection}[1]{\setcounter{equation}{0}\section{#1}}

\newcommand{\ket}[1]{\ensuremath{\mbox{\normalsize $| #1\rangle$}}}
\newcommand{\sgn}{{\rm sgn}}
\begin{document}
\begin{titlepage}
\vskip 0.5cm
\begin{flushright}
DCPT-04/31\\
SPhT-T04/050\\
{\tt hep-th/0410013}\\
\end{flushright}
\vskip 1.5cm
\begin{center}
{\Large{\bf
Beyond the WKB approximation in\\[5pt] 
$\PT$-symmetric quantum mechanics
}}
\end{center}
\vskip 1.0cm
\centerline{Patrick Dorey$^{1,2}$, Adam Millican-Slater$^1$
and Roberto Tateo$^3$}
\vskip 0.7cm
\centerline{${}^1$\sl\small Dept.~of Mathematical Sciences,
University of Durham,}
\centerline{\sl\small  Durham DH1 3LE, United Kingdom\,}
\vskip 0.3cm
\centerline{${}^2$\sl\small Service de Physique Th{\'e}orique,
CEA-Saclay,}
\centerline{\sl\small F-91191 Gif-sur-Yvette Cedex, France}
\vskip 0.3cm
\centerline{${}^{3}$\sl\small Dip. di Fisica Teorica and INFN,
Universit\`a di Torino,}
\centerline{\sl\small Via P. Giuria 1, 10125 Torino, Italy}
\vskip 0.4cm
\centerline{E-mails:}
\centerline{\tt p.e.dorey@durham.ac.uk}
\centerline{\tt adam.millican-slater@durham.ac.uk}
\centerline{\tt tateo@to.infn.it}

\vskip 1.25cm
\begin{abstract}
\noindent
The mergings of energy levels associated with the breaking of $\PT$
symmetry in the model of Bender and Boettcher, and in its 
generalisation to incorporate a centrifugal term, are analysed in 
detail. Even though conventional WKB techniques fail, it is shown 
how the ODE/IM correspondence can be used to obtain a systematic 
approximation scheme which captures all previously-observed features.
Nonperturbative effects turn out to play a crucial role, governing the
behaviour of almost all levels once the symmetry-breaking transition
has been passed. In addition, a novel treatment of the  radial 
Schr\"odinger equation is used to recover the values of local and 
non-local conserved charges in the related integrable quantum field 
theories, without any need for resummation even when the angular 
momentum is nonzero.
\end{abstract}
\end{titlepage}
\setcounter{footnote}{0}
\def\thefootnote{\fnsymbol{footnote}}
%
%
\resection{Introduction}
\label{intr}
{}Following a paper by Bender and Boettcher \cite{Bender:1998ke},
itself inspired by a conjecture of Bessis and
Zinn-Justin \cite{BZJ}, 
the subject of $\PT$-symmetric quantum 
mechanics has attracted growing interest. (A couple of
earlier papers exploring related themes 
are \cite{CGM1980,BG1993}, while
\cite{Bender:1998gh}--\cite{Kleefeld:2004qs}
provide a small sample of subsequent work.)
Many studies have discussed general
interpretational questions, but there are also mathematical 
issues to be addressed.
One such
is the topic of this paper: the pattern of $\PT$-symmetry  breaking
in the original model of Bender and Boettcher 
\cite{Bender:1998ke,Bender:1998gh},
and in the generalisation of their model that was introduced in
\cite{Dorey:1999uk}. Numerical work has
demonstrated an intriguing pattern of merging energy levels associated
with the symmetry-breaking transition, but a complete analytic understanding
has proved elusive, partly because standard WKB techniques break
down in the regime where the symmetry-breaking occurs~\cite{Bender:1998ke}. 
In this paper we show how this problem can be overcome, using a
recently-discovered link with the theory of integrable quantum field theory
\cite{Dorey:1998pt} 
(the `ODE-IM correspondence')
to develop an approximation scheme for the energy
levels which captures all features of the previously-observed
behaviour in a controlled manner.

The remainder of this paper is organised as follows.  
Section~\ref{PTpheno} describes the symmetry-breaking 
transition in more detail, 
while section~\ref{thpt} discusses earlier theoretical treatments.
The relevant connections with integrable quantum field theory are
recalled in section~\ref{corresp}, 
and then applied to the problem at hand in 
section~\ref{Tasymp}. Section~\ref{Qasymp} then shows 
how a novel treatment of the radial anharmonic oscillator can
be used to recover some of the results used earlier,
calculations which may be of independent interest for the ODE-IM 
correspondence since they relate to the values of 
conserved charges in certain integrable quantum
field theories.
Finally, section~\ref{concl} contains our conclusions.

\resection{Some $\PT$ phenomenology}
\label{PTpheno}
In the early 1990s, Bessis and Zinn-Justin\,\cite{BZJ}
conjectured that the
non-Hermitian Hamiltonian
\eq
\CH=p^2+i\,x^3
\en
should have a real and positive spectrum, if the boundary condition
$\psi\in L^2(\R)$ is imposed on the position-space wavefunctions
$\psi(x)$. This initially-surprising proposal was motivated
by considerations of the quantum field theory of the Lee-Yang model;
it was subsequently put into a more directly
quantum-mechanical
context by Bender
and Boettcher 
\cite{Bender:1998ke}, who also suggested that
the spectral properties
of the one-parameter family of Hamiltonians
\eq
\CH\php_M=p^2-(ix)^{2M}
\label{bbham}
\en
might be of interest, with
$M$ a positive real number. Varying $M$ from
 $3/2$ to $1$ interpolates between $\CH_{3/2}$\,, the Bessis--Zinn-Justin
Hamiltonian, and $\CH_1$\,, the much more familiar Hamiltonian of
the simple harmonic oscillator.
For $M\ge 2$, analytic continuation (in $M$) of the eigenvalue problem
requires boundary conditions to be imposed on a suitably-chosen contour in
the complex plane, away from the real axis \cite{Bender:1998ke}; given this,
Bender and Boettcher found that the conjecture of Bessis and Zinn-Justin
could be strengthened to the statement that the spectrum of $\CH\php_M$
is real and positive for all $M\ge 1$.
This reality property is associated with, but not completely
explained by, the so-called
$\PT$ symmetry of the corresponding spectral problems.
In fact, it is only recently that a complete proof of the
Bender-Boettcher conjecture has been given
\cite{Dorey:2001uw}.

{~}\vskip -15pt
\[
\!\!\!\!\!
\begin{array}{cc}
\includegraphics[width=0.44\linewidth]{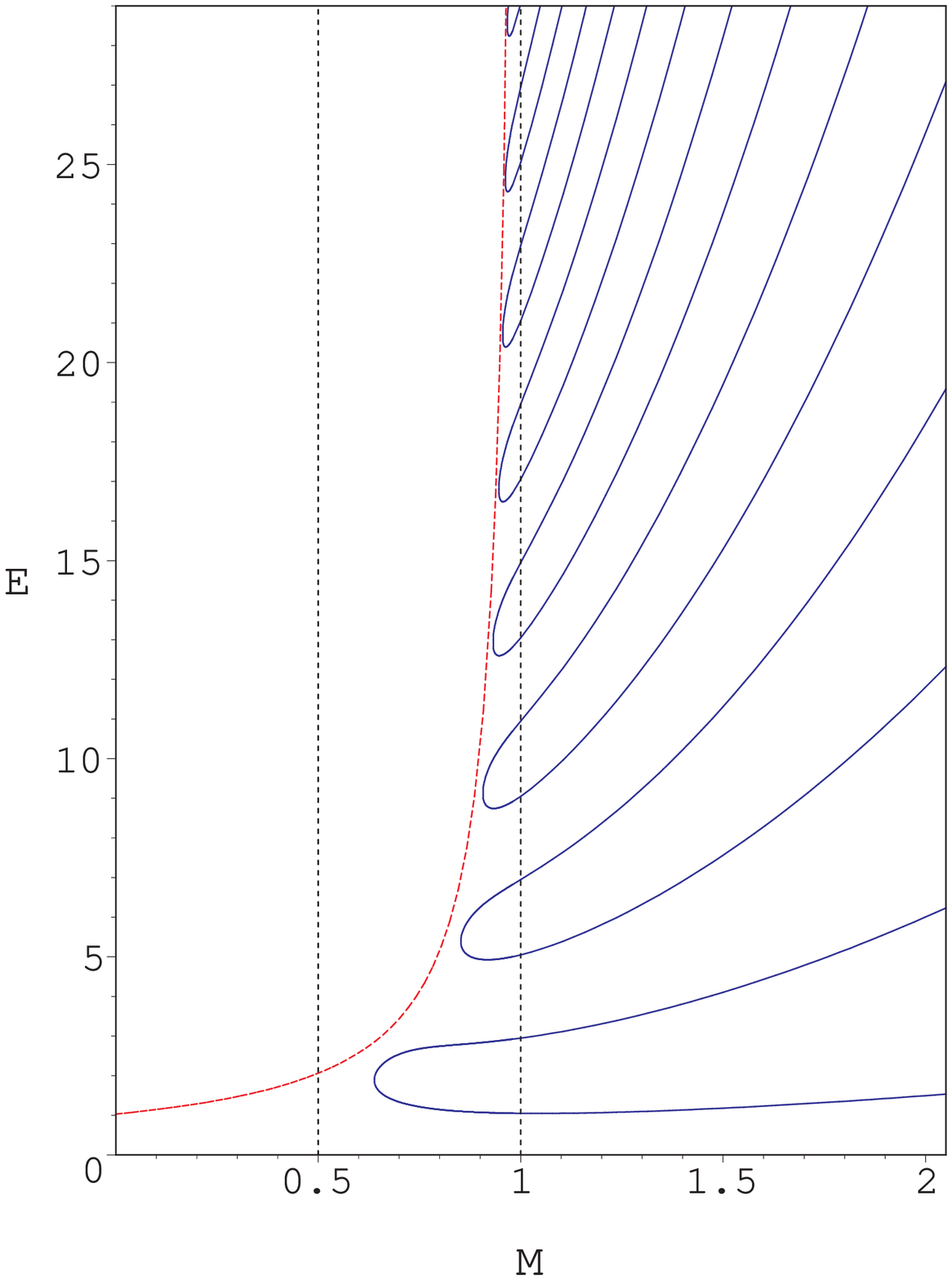}~~~&~~~~%
\includegraphics[width=0.44\linewidth]{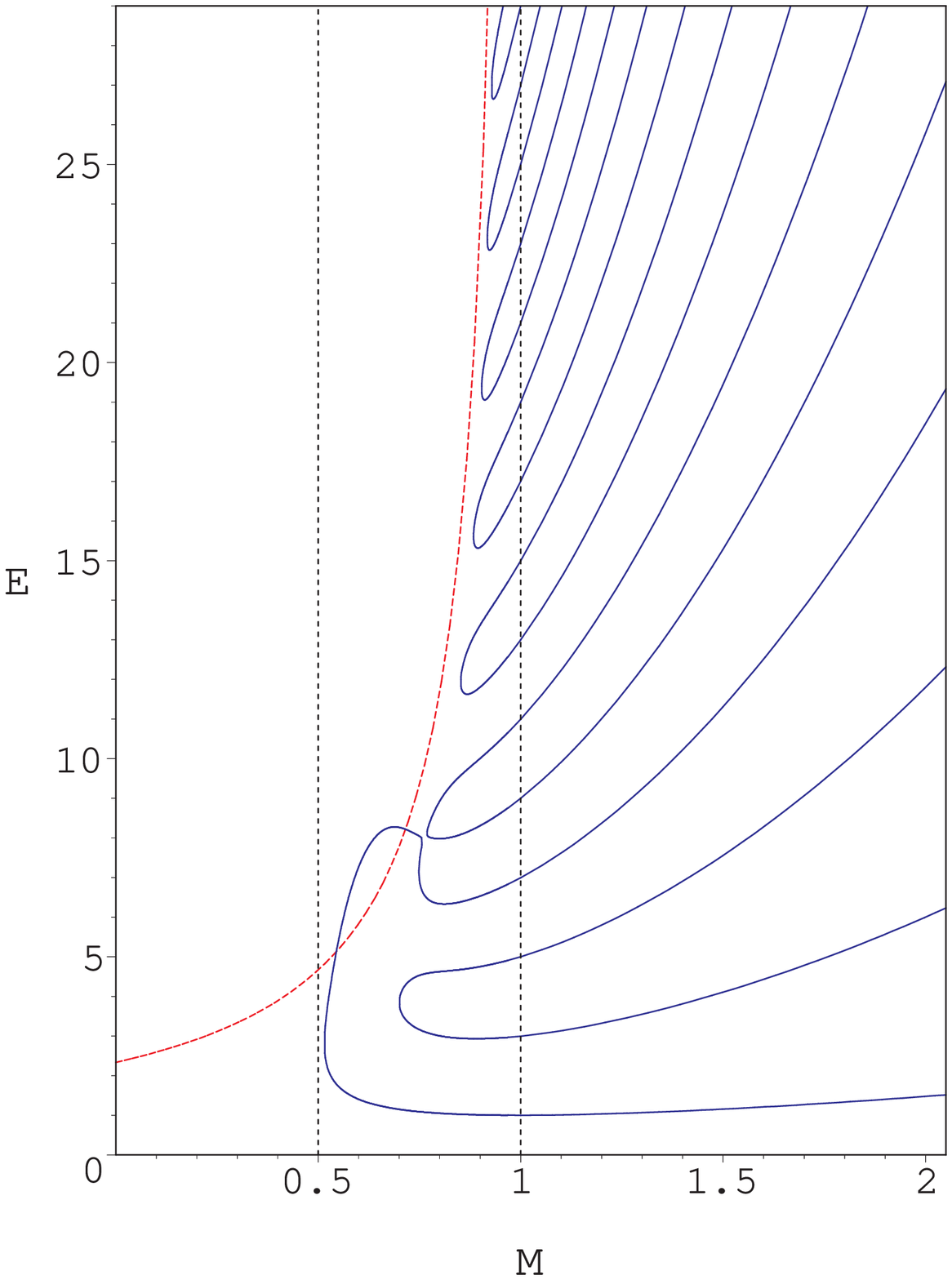}\\[2pt]
\parbox{0.33\linewidth}{~~~~~~\small\protect\ref{numerics}a:
$l=-0.025$}~~~~&~~~~
\parbox{0.33\linewidth}{~~~~~~\small\protect\ref{numerics}b:
$l=-0.001$}\\[24pt]
\includegraphics[width=0.44\linewidth]{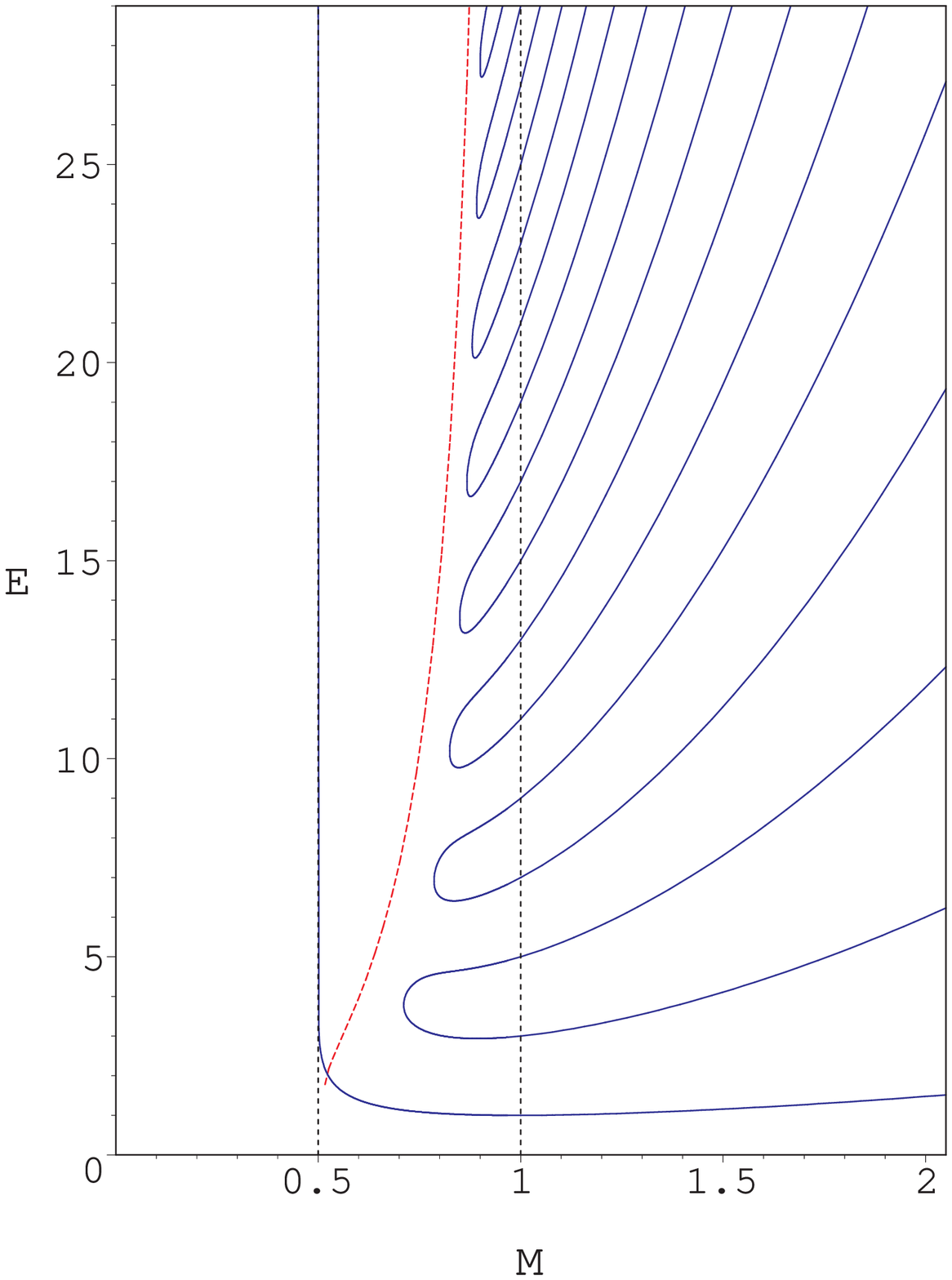}~~~&~~~~%
\includegraphics[width=0.44\linewidth]{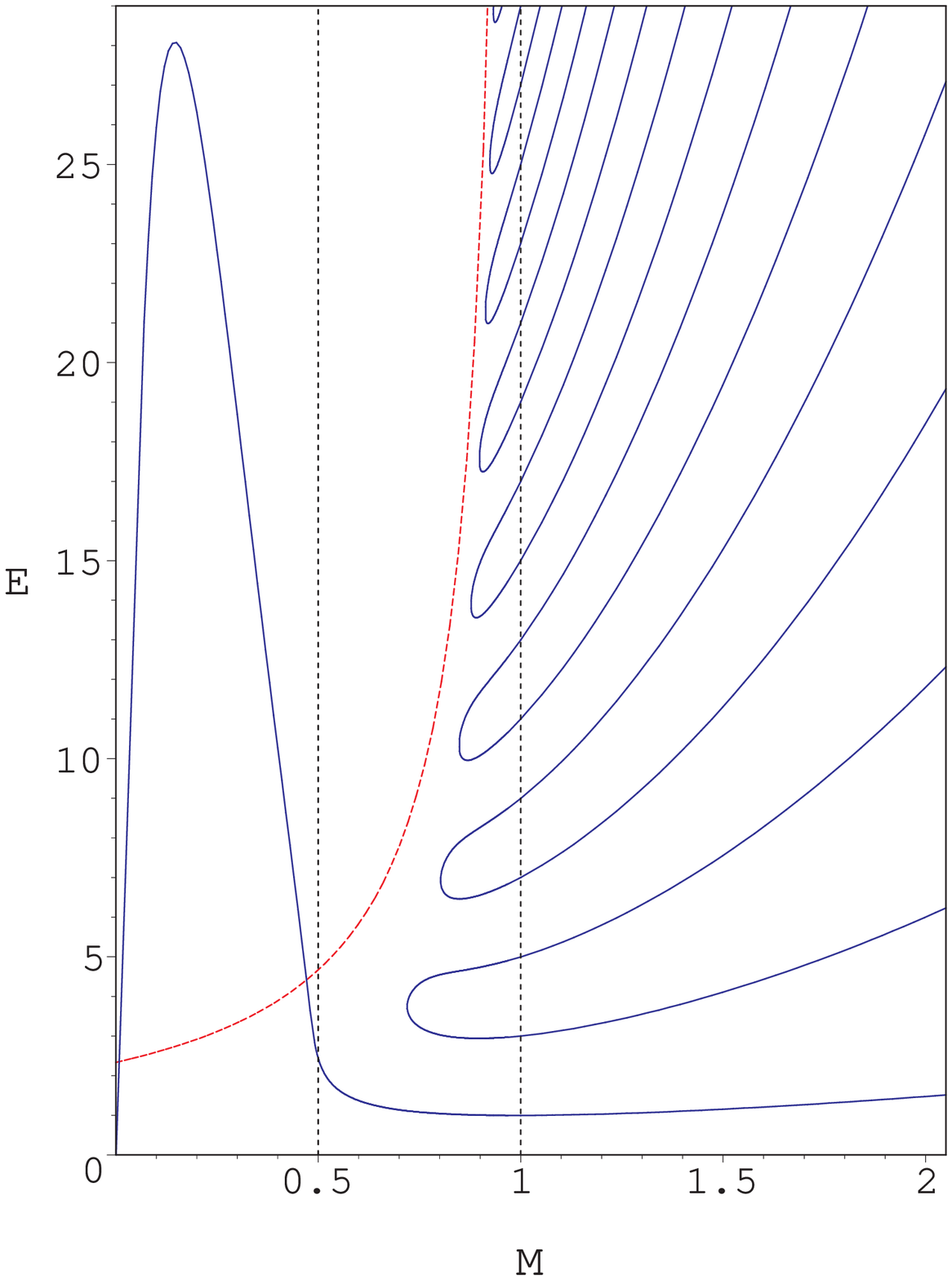}\\[2pt]
\parbox{0.33\linewidth}{~~~~~~\small\protect\ref{numerics}c: $l=0$}~~~~&~~~~
\parbox{0.33\linewidth}{~~~~~~\small\protect\ref{numerics}d: $l=0.001$}\\[12pt]
\multicolumn{2}{l}%
{~~
\parbox{.99\linewidth}{{\small\noindent%
Figure \protect\ref{numerics}: Real eigenvalues of
$p^2-(ix)^{2M}+l(l{+}1)/x^2$ as functions of $M$,
for various values of $l$. The (curved) dotted lines show the asymptotic
locations of the level-mergings, to be discussed in 
section~\protect\ref{Tasymp} below.%
}
}
}
\end{array}
\refstepcounter{figure}\label{numerics}
\]

Most interesting, though, is the behaviour of the spectrum as
$M$ falls below $1$, shown in figure~\ref{numerics}c.
Numerical results
indicate that
infinitely-many eigenvalues collide in pairs and become
complex, leaving a finite number of real
eigenvalues\,\cite{Bender:1998ke}.
As $M$ decreases
further, these successively pair off and become complex until finally
the second and third levels merge and only the ground state
remains real, which itself diverges to infinity as $M\to \frac{1}{2}^+$.
(For $M=\frac{1}{2}$\,, the spectrum of (\ref{bbham})
is null, as can be seen by shifting $ix$ to $ix-E$  and
solving the differential equation using the Airy function.) 

The mysterious nature of the transition at $M=1$ is
highlighted by a further generalisation of Bessis and
Zinn-Justin's Hamiltonian, to incorporate an
additional `angular momentum' term  \cite{Dorey:1999uk}:
\eq
\CH\php_{M,l}=p^2-(ix)^{2M}+\frac{l(l{+}1)}{x^2}\,.
\label{ush}
\en
Here
$l$ is a real parameter, which can be assumed no smaller than $-1/2$,
since the problem is unchanged by the replacement of
 $l$ by $-1-l$\,. For $l\neq 0$,
the contour on which the wavefunction is defined should be distorted
below the singularity at $x=0$\,, in addition to any distortions away from
the real axis required for $M\ge 2$.
The spectrum of $\CH_{M,l}$ is then real for all
$M\ge 1$, and positive if in
addition $l<M/2$ \cite{Dorey:2001uw}. As before, there is a
transition at $M=1$, with infinitely-many levels becoming complex. However,
as illustrated in figure \ref{numerics}a,
for sufficiently-negative values of
$l$ there is a remarkable change in the way that the
remaining real levels pair off:
the parity is reversed, so that the second level is paired not with
the third level, but rather with the ground state, and so
on up the spectrum. This reversed pairing holds true for sufficiently-high
levels for all
negative $l$, but as $l$ gets closer to
zero the level with which the ground state
is paired moves up through the spectrum, leaving the
connectivity of the original Bender-Boettcher problem
in its wake and allowing for a continuous transition to
the previous behaviour at $l=0$.
This is described at greater length in \cite{Dorey:1999uk}, and
is best understood by looking at figures~\ref{numerics}a --
\ref{numerics}d, or
at figure~2 of \cite{Dorey:1999uk};
the challenge is to obtain an analytic understanding of why
it occurs.

\resection{Some $\PT$ Theory}
\label{thpt}
A first insight into the phenomena described in the last section
comes from the observation that, while $\CH_M$ is not Hermitian
in any simple sense, it {\em is}\/ invariant under the combined action of the
operators $\CP$ and $\CT$ 
\cite{Bender:1998ke}, where $\CP$ is parity and $\CT$ time
reversal, acting on Schr\"odinger potentials $V(x)$ as
\eq
\CP V(x) \CP^{-1}=V(-x^*)~~,~~~~
\CT V(x) \CT^{-1}=V(x)^*~.
\en
(The complex conjugation in the definition of parity ensures that the deformed
contours required for $M\ge 2$ are mapped onto themselves, but is otherwise
unimportant.) As shown in
\cite{Bender:1998gh}, $\PT$ invariance implies that
eigenvalues are either real, or come in complex-conjugate pairs, much like
the roots of a real polynomial (see also \cite{Bender:2002yp}).
Real eigenvalues correspond to wavefunctions symmetrical under $\PT$,
complex eigenvalues to a spontaneous breaking of this symmetry.
In contrast to Hermiticty,
on its own $\PT$ symmetry is not enough to prove reality;
but it does mean
that if a level is to become complex, then it must pair off with some
other real level first.

To deal with the transition at $M=1$,
Bender, Boettcher and Meisinger \cite{Bender:1998gh}
used a basis of
the exactly-known eigenfunctions $\ket{n}$ of $\CH_1$,
the simple Harmonic oscillator:
\eq
\CH_1\ket{n}=(2n{+}1)\,\ket{n}~,\quad n=0,1,\dots
\en
and approximated the Hamiltonian for small $\epsilon := 2(M{-}1)$ as
\eq
\CH_{1+\epsilon/2}=p^2+x^2+\epsilon x^2\left[
\ln|x|+\fract{i\pi}2\sgn(x)\right] +O(\epsilon^2)\,.
\en
Truncating the approximate Hamiltonian to the two-dimensional
subspace spanned by $\ket{2n{-}1}$ and $\ket{2n}$,
for large $n$ and small $\epsilon$
the matrix elements are \cite{Bender:1998gh}
\eq
\CH_{\rm trunc}\approx
\left(
\begin{array}{cc}
a(2n{-}1)& ib(2n) \\[3pt]
ib(2n) & a(2n)
\end{array}
\right)
\en
where
\eq
a(n)=2n+1+\frac{\epsilon\, n}{2}\ln(n)\quad,\quad
b(n)=\frac{4}{3}\,\epsilon\, n~.
\en
Diagonalising $\CH_{\rm trunc}$ yields the prediction that
the energy levels of $\CH_{1+\epsilon/2}$ should join and
become complex for
\eq
|\epsilon|\sim \frac{3}{8n}~,
\label{trancas}
\en
matching the numerical observation that the points at
which levels merge approach the line $M=1$ as the level $n$ tends to
infinity.

However, the agreement of this result with the numerical data
is only qualitative, and even at this level,
there are problems. Firstly,
the approximation also predicts a merging of levels for {\em positive}\/
values of $\epsilon$, which is clearly wrong; and secondly, if
one instead truncates to the subspace spanned by the levels $\ket{2n}$
and $\ket{2n{+}1}$, then the same approximation predicts that these
two levels should pinch off. As underlined by the effect of
the angular-momentum
term, it is by no means obvious
{\em a priori}\/
how the pairing of
levels should go, and it
is reasonable to demand that any full
understanding of the transition at $M=1$
should include a robust (and correct!) prediction on this point.
Figures \ref{truncfail}a and \ref{truncfail}b
illustrate the situation.

\[
\begin{array}{ll}
\includegraphics[width=0.35\linewidth]{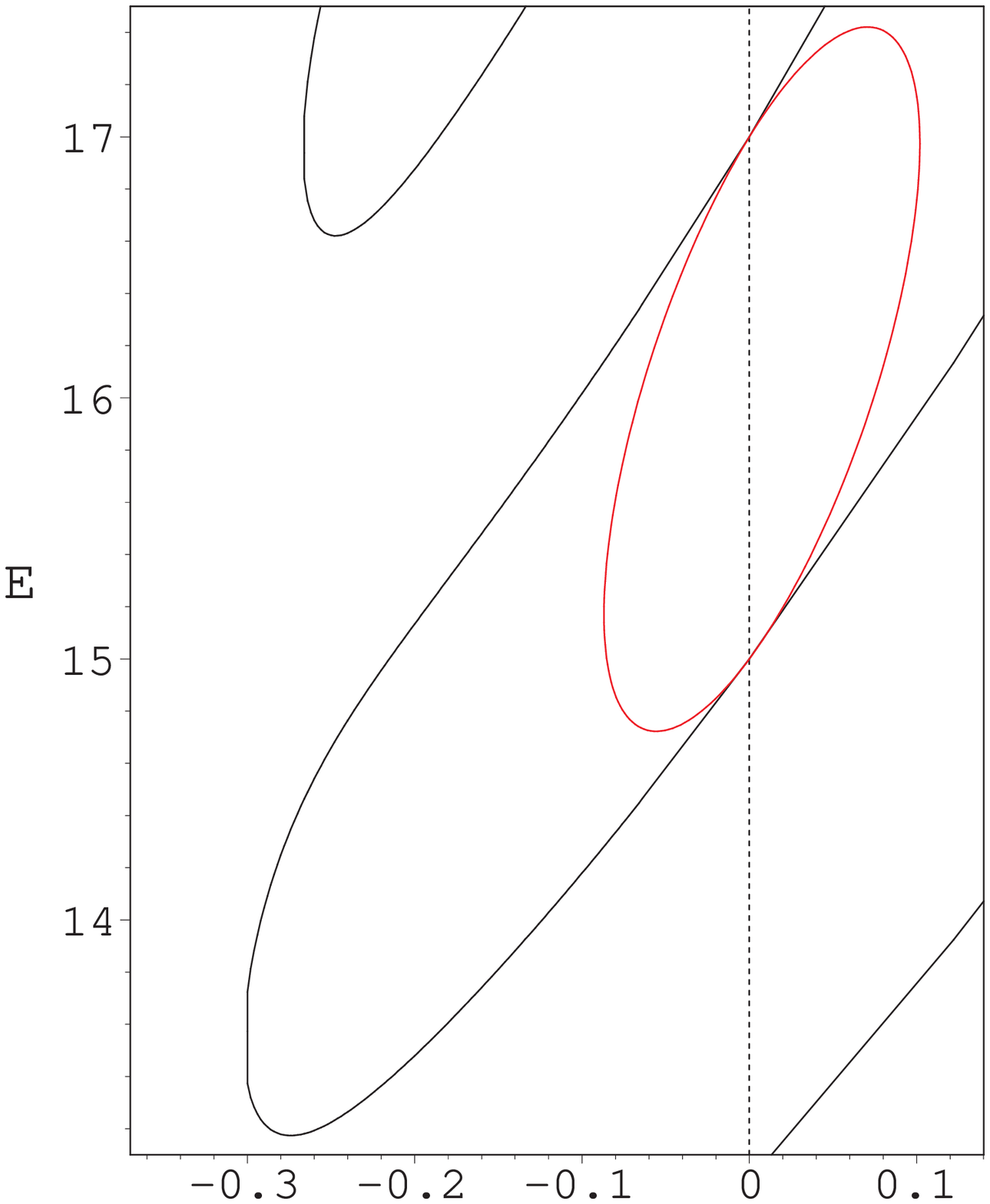}~~&~~
\includegraphics[width=0.35\linewidth]{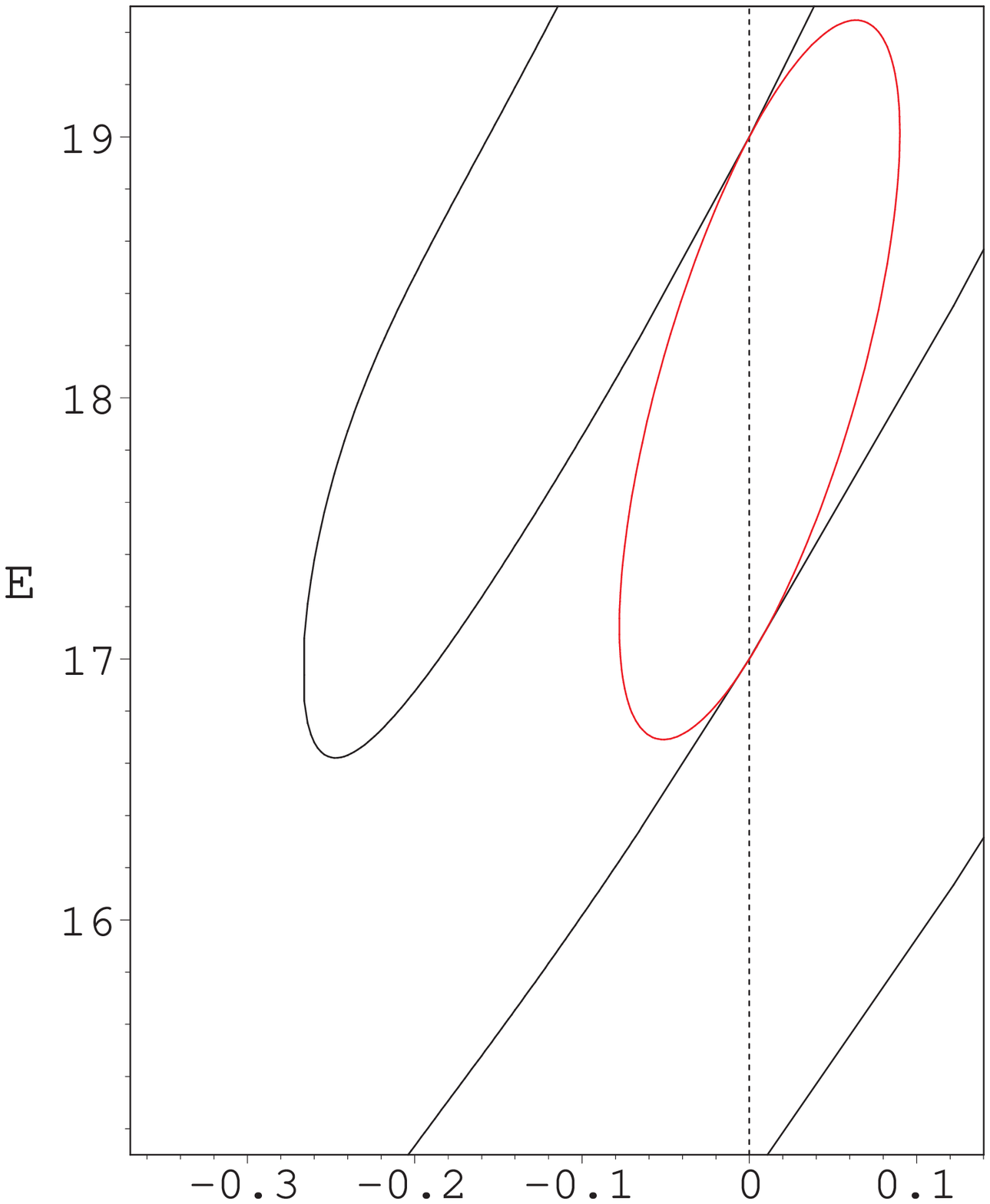}
\\[3pt]
\parbox{0.381\linewidth}{\small Figure \protect\ref{truncfail}a:
Truncation to levels 8 and 9, compared with numerical data}~~~~~~~&~~~
\parbox{0.381\linewidth}{\small Figure \protect\ref{truncfail}b:
Truncation to levels 9 and 10, compared with numerical data}
\end{array}
\refstepcounter{figure}\label{truncfail}
\]

The discussion of \cite{Bender:1998gh} concerned
the zero angular-momentum case.
Since this is on the boundary between the two sorts of
level connectivity, one might hope for a clearer signal from the
truncated Hamiltonian at $l\neq 0$, where the matrix elements
can be given in terms of hypergeometric functions.
However, we have found that
the problems described above persist. The approximation
(both for zero and for non-zero angular momentum) does improve if
the dimension of the truncated subspace is increased, but this rapidly
reverts to a numerical treatment, and any analytical understanding
is lost. Alternatively, it might be hoped
that the full set of matrix elements
in the harmonic oscillator basis would simplify sufficiently in a
suitable large-$n$, small-$\epsilon$ limit that a precise prediction
could be made. While this seems a technically-interesting avenue
to explore, it is a delicate exercise to extract the
relevant asymptotics, and we have yet to make useful progress
in this direction.

In more standard
quantum-mechanical problems, methods such as truncation are
not necessary when discussing the asymptotic behaviour of the high-lying
levels, as the WKB approximation can be used. Indeed,
this method gives very accurate results for the current problem
when $M\ge 1$ \cite{Bender:1998ke}. However, the direct application of WKB
{\em fails}\/ for $M<1$ 
\cite{Bender:1998ke}: the path along which the phase-integral
quantization condition should be taken crosses a cut and ceases to join
the relevant two turning points. A na\"ive analytic continuation
of the WKB results
for $M\ge 1$ to $M<1$
also fails, producing energy levels which do not even become
complex.

In the following we shall show how these problems can
be avoided, by developing an approximation scheme using
the so-called ODE/IM
correspondence, which links the Bender-Boettcher problem and its
generalisations to a set of integrable quantum field theories associated
with the sine-Gordon model. For $M\ge 1$ the WKB approximation is
recovered, but for $M<1$ the results are significantly different.
In preparation, the next section gives a brief review of
the correspondence.

\resection{More $\PT$ Theory: the ODE/IM correspondence}
\label{corresp}
In its simplest form, the ODE/IM correspondence links the spectral
properties of certain Schr\"odinger equations to functions which
appear in the study of integrable quantum field theories in 1+1
dimensions. The phenomenon was first observed in
\cite{Dorey:1998pt}, for Schro\"odinger problems with homogeneous
potentials
$V(x)=x^{2M}$. The correspondence was generalised to incorporate an
additional angular-momentum like term in
\cite{Bazhanov:1998wj} and a more general potential in \cite{Sc},
while its relevance to the non-Hermitian spectral problems of
$\PT$-symmetric quantum mechanics was found in
\cite{Dorey:1999uk}.
A longer review of the subject is given in
\cite{Dorey:2000kq}, and here we shall just summarise the key results,
largely following \cite{Dorey:1999uk}.

Let $\{E_i\}$ be the set of eigenvalues of
$\CH\php_{M,l}$\,, defined in (\ref{ush}), and let $\{e_j\}$ be the
eigenvalues of a similar-looking but Hermitian problem
\begin{equation}
\left(-\frac{d^2}{dx^2}+x^{2M}+\frac{l(l{+}1)}{x^2}\right)\psi=E\psi
\label{hprob}
\end{equation}
with so-called `radial' boundary conditions
$\psi(x)= O(x^{l+1})$ as $x\to 0$\,,
$\psi(x)\to 0$ as $x\to\infty$.
(Correspondingly, the earlier, $\PT$-symmetric problems are sometimes
said to have `lateral' boundary conditions.)
A pair of spectral determinants for these two problems, $T(E)$ and $Q(E)$,
can be defined as follows. For any $E$, not necessarily
an eigenvalue of either problem, let $y(x,E,l)$ be the unique solution
to (\ref{hprob}) which has the same asymptotic as $x\to +\infty$
on the real axis as
\begin{equation}
y^{\rm WKB}(x,E) :=
\frac{1}{\sqrt{2i}}\,P(x)^{-1/4}\,\exp\left(-\!\int_0^x\!\sqrt{P(t)}\,dt\right)
\label{wkbdef}
\end{equation}
where $P(x)=x^{2M}-E$. (Note,
so long as $M$ is positive, $l$
is not involved in the specification of this asymptotic.)
To give some examples:
\begin{eqnarray}
\mbox{$M>1$\,:\qquad}
y(x,E,l) &\sim &
\frac{ 1 }{\sqrt{2i}}\,
x^{-M/2}\,
\exp\left(-\fract{1}{M{+}1}\,x^{M{+}1}\right)
\\
\mbox{$M=1$\,:\qquad}
y(x,E,l) &\sim &
\frac{1}{\sqrt{2i}}\, x^{-1/2+E/2}\,
\exp\left(-\fract{1}{2}\,x^{2}\right)
\\
\mbox{$1>M>\frac{1}{3}$\,:\qquad}
y(x,E,l) &\sim &
\frac{1}{\sqrt{2i}}\,x^{-M/2}\, \exp\left(-\fract{1}{M{+}1}\,x^{M{+}1}+
 \fract{E}{2{-}2M}\,x^{1-M}\right)\qquad
\end{eqnarray}
More generally, the explicit asymptotic of $y(x,E,l)$ changes whenever
$M=1/(2m{-}1)$ with $m$ a positive integer, as is easily seen from
(\ref{wkbdef}). We also define $\psi(x,E,l)$ to be the solution
to (\ref{hprob}) which behaves as $x\to 0$ as
\begin{equation}
\psi(x,E,l)\sim x^{l+1}+O(x^{l+3})\,.
\end{equation}
Here we continue to take $\Re e\,l \ge -1/2$\,, making the definition
unique; if desired, a second
solution $\psi(x,E,-1-l)$ to (\ref{hprob}) can be uniquely specified
by analytic continuation in $l$.

Denoting the Wronskian $fg'-f'g$ of two functions $f(x)$ and $g(x)$ by
$W[f,g]$\,, we now define $T$ and $Q$ by
\begin{eqnarray}
T(E)&:=&W[y(\omega x,\omega^{-2}E,l),y(\omega^{-1}x,\omega^2E,l)]
\\[3pt]
Q(E)&:=&W[y(x,E,l),\psi(x,E,l)]
\end{eqnarray}
where
\begin{equation}
\omega=e^{i\pi/(M{+}1)}\,.
\label{omegadef}
\end{equation}
Then it is straightforwardly shown
\cite{Dorey:1999uk} that the zeroes of $T$ are exactly the
points $\{-E_i\}$, and the zeroes of $Q$ are the points $\{e_j\}$\,,
so these two functions are indeed spectral determinants.
Furthermore they obey the following functional relation:
\begin{equation}
T(E)\,Q(E)=\omega^{-l-1/2}\,Q(\omega^{-2}E)+\omega^{l+1/2}\,Q(\omega^{2}E)
\label{tq}
\end{equation}
which is the same as Baxter's T-Q relation \cite{baxterbook} from the
theory of integrable lattice models, in the
form which arises in
connection with the
quantum field theory of the massless sine-Gordon model
\cite{Bazhanov:1996dr}. (Strictly speaking, for $l\neq -1/2$\,,
$Q$ as defined here
corresponds to the function denoted $A$ in \cite{Bazhanov:1996dr}.)
At the points $M=1/(2m{-}1)$ where the asymptotic form of $y$ changes,
a correction term appears in (\ref{tq})
\cite{Bazhanov:1996dr,Dorey:1999uk},
but the
only case potentially relevant in the following is $M=m=1$, to which
we shall return briefly below.

Now define
\begin{equation}
a(E):=\omega^{2l+1}\,\frac{Q(\omega^2E)}{Q(\omega^{-2}E)}~.
\label{adef}
\end{equation}
By (\ref{tq}),
$a(E)=-1$ precisely at the zeroes of $T(E)$ and $Q(E)$,
that is at the set of points
$\{-E_i\}\cup\{e_j\}$\,\footnote{Note, exceptionally,
the zeroes of the LHS of (\ref{tq}) might coincide with
simultaneous zeroes of the factors on the RHS, but this
will not concern us below.}. Furthermore,
provided the $\{e_j\}$ lie on the positive real axis and the
$\{-E_i\}$ lie away from it,
$a(E)$ can be found by solving a nonlinear integral equation,
as follows \cite{KBP,Destri:1994bv,Bazhanov:1996dr}.
Trade $E$ for a `rapidity' $\theta(E)$, such that
\begin{equation}
E^{(M+1)/(2M)}=re^{\theta}.
\label{thdef}
\end{equation}
Then, for $|\Im m\,\theta|<\min(\pi,\pi/M)$\,, the `counting function'
\begin{equation}
f(\theta):=\log a(E(\theta))
\label{fdef}
\end{equation}
solves
\begin{eqnarray}
f(\theta)&=&i\pi(l{+}\fract{1}{2})-imre^{\theta}
+\int_{C_1}\varphi(\theta{-}\theta')\log(1+e^{f(\theta')})\,d\theta'
\nn\\[3pt]
&&\qquad\qquad\qquad\qquad-
\int_{C_2}\varphi(\theta{-}\theta')\log(1+e^{-f(\theta')})\,d\theta'
\label{nlie}
\end{eqnarray}
where
\begin{equation}
 m=
\sqrt{\pi}\,
\frac{\Gamma(1+\frac{1}{2M})}{\Gamma(\frac{3}{2}+\frac{1}{2M})}
=\frac{1}{M}B(\fract{3}{2},\fract{1}{2M})
\label{mdef}
\end{equation}
and
\begin{equation}
B(p,q)=\frac{\Gamma(p)\Gamma(q)}{\Gamma(p+q)}=
\int_0^1t^{p-1}(1-t)^{q-1}dt
\label{betaf}
\end{equation}
is Euler's beta function.
The integration contours $C_1$ and $C_2$ in (\ref{nlie})
run from $-\infty$ to $+\infty$
just below and just above the real axis, close enough that all
of the points $\{\theta(e_j)\}$, and none of
the points $\{\theta(-E_i)\}$,
lie between $C_1$ and $C_2$\,,,
and the kernel function $\varphi(\theta)$ is
\begin{equation}
\varphi(\theta)=\int_{-\infty}^{\infty}e^{ik\theta}
\frac{\sinh\left(\frac{\pi}{2}\frac{1{-}M}{M}k\right)}%
{2\cosh\left(\frac{\pi}{2}k\right)
\sinh\left(\frac{\pi}{2M}k\right)}\,
\frac{dk}{2\pi}~.
\label{kerdef}
\end{equation}
The value of the normalisation factor $r$ is arbitrary, but to match the
conventions of \cite{Bazhanov:1996dr,Dorey:1999uk} one should take
\begin{equation}
r= (2M{+}2)\,\Gamma\!\left(\frac{M}{M{+}1}\right)^{(M{+}1)/M}.
\end{equation}

The nonlinear integral equation (\ref{nlie}) must be modified if any
$\theta(e_j)$ moves outside the contours $C_1$ and $C_2$
\cite{Bazhanov:1996dr}, or if any $\theta(-E_i)$
moves inside them \cite{Fioravanti:1996rz}.
However
these possibilities do not arise in the current context, at least for
small values of $l$:
for $M\ge 1$ and $|2l{+}1|<M{+}1$
it can be shown that all of the $e_j$ lie on the positive
real axis of the complex $E$ plane,
and all of the $-E_i$ on the negative real axis
(see \cite{Dorey:1999uk,Dorey:2001uw}). For $M<1$, the arguments
showing that the
$\{e_j\}$ lie on the positive real axis continue to hold, but, as we
already saw, the $\{-E_j\}$ do move away from the negative real axis
and become complex. However, our numerical results show clearly that
they never become near enough to the real axis to upset the arguments
below.

\resection{Asymptotics from the integral equation}
\label{Tasymp}
Equation (\ref{nlie}) provides an
effective way to solve the radial spectral problem (\ref{hprob})
\cite{Dorey:1998pt}\,:
given that the zeroes of $T$ all lie away from the positive real $E$
axis, any point on the real $\theta$ axis at
which $f(\theta)=(2n{+}1)\pi i$ for some $n\in\Z$
must correspond to a zero of $Q$.
The values of $f(\theta)$ on $C_1$ and $C_2$ can be obtained
numerically from (\ref{nlie}) by a simple iterative procedure, after
which the same equation can be used to obtain $f(\theta)$ on the real
axis, allowing those points at which
$f(\theta)=(2n{+}1)\pi i$ to be located with high accuracy.
The leading approximation for $f(\theta)$, found by dropping
the integrals
from (\ref{nlie}), yields the usual WKB result; higher corrections
can be obtained using the asymptotic expansion for Q(E) in terms of
local and non-local charges given
in~\cite{Bazhanov:1996dr,Bazhanov:1998za}.

The same numerical approach was used in
\cite{Dorey:1999uk} to analyse the non-Hermitian spectral problems
based on (\ref{ush}); it was also used to generate
figure~\ref{numerics} above. However an analysis of the
implications for the asymptotic behaviour of the energy levels was not
given. Compared to the Hermitian case, there are a number of
subtleties, which we now discuss.

The first point, which was already taken into account
in the numerical work described
in \cite{Dorey:1999uk}, is that the zeroes of $T(E)$ are
either on or near to the {\em negative}\/ real $E$-axis,
which is the line $\Im m\,\theta
=\pi (M{+}1)/(2M)$ on the complex $\theta$-plane. Since
$(M{+}1)/(2M)>\min(1,1/M)$ for all $M\neq 1$, this means that the
relevant values of $f(\theta)$ lie outside the strip described by
(\ref{nlie}).
Instead, the so-called `second determination' must be used
\cite{Destri:1997yz}.
This arises
because the kernel $\varphi(\theta)$ has, amongst others, poles
at $\theta=\pm i\pi$ and $\pm i\pi/M$, which add residue terms to the
analytic continuation
(\ref{nlie}) outside the strip
$|\Im m\,\theta|<\min(\pi,\pi/M)$\,.
(See~\cite{Dorey:1996re}
for related discussions in the
context of the thermodynamic Bethe ansatz.)

For $M>1$ (sometimes called the attractive, or semiclassical, regime
in the quantum field theory context) the poles in
$\varphi(\theta)$ nearest to the real axis are at $\theta=\pm i\pi/M$,
and have residue $-\frac{1}{2\pi i}$.
Increasing $\Im m\,\theta$ past $\pi/M$\,,
the pole at $\theta=-i\pi/M$ crosses
$C_1$ and then $C_2$, generating
a term $-f(\theta{-}i\pi/M)$ which should be
added to the right-hand side of (\ref{nlie}) in order to give a
correct representation of $f(\theta)$ for $\Im m\,\theta$ just greater
than $\pi/M$. Using the original
representation (\ref{nlie}) to rewrite the extra term,
the resulting expression is
\begin{eqnarray}
f(\theta)&=&-i(1{-}e^{-i\pi/M})\,mre^{\theta}
+\int_{C_1}\varphi_{II}(\theta{-}\theta')\log(1+e^{f(\theta')})\,d\theta'
\nn\\[3pt]
&&\qquad\qquad\qquad\qquad\qquad -
\int_{C_2}\varphi_{II}(\theta{-}\theta')\log(1+e^{-f(\theta')})\,d\theta'
\qquad
\label{attrnlie}
\end{eqnarray}
where
\begin{equation}
\varphi_{II}(\theta)=\varphi(\theta)-\varphi(\theta{-}i\pi/M)=
\frac{2i\,\cos(\frac{\pi}{2M})\sinh(\theta{-}\frac{i\pi}{2M})}%
{\pi(\cosh(2\theta{-}\frac{i\pi}{M})-\cos(\frac{\pi}{M}))}~.
\end{equation}
An examination of the poles in $\varphi_{II}(\theta)$ shows that
(\ref{attrnlie}) holds for
$\pi/M<\Im m\,\theta<\pi$, which includes the neighbourhood of
the line $\Im m\,\theta=\pi(M{+}1)/(2M)$ relevant to
the hunt for the zeroes of $T(E)$. (For simplicity we have supposed
that $C_1$ and $C_2$ are infinitesimally close to the real axis;
otherwise
an `intermediate determination' is also required, applying when
the pole in $\varphi(\theta)$ has crossed $C_1$ but not $C_2$\,.)

For $M<1$ (the `repulsive regime' in quantum field theory
language) the nearest poles are instead at $\theta=\pm i\pi$, and for
$\Im m\,\theta>\pi$ the extra term to be added to the right-hand side
of (\ref{nlie}) is equal to $+f(\theta{-}i\pi)$. Rewriting using
the original representation as before, one finds
\begin{eqnarray}
f(\theta)&=&2i\pi(l{+}\fract{1}{2})
+\int_{C_1}\varphi_{II}(\theta{-}\theta')\log(1+e^{f(\theta')})\,d\theta'
\nn\\[3pt]
&&\qquad\qquad\qquad\qquad-
\int_{C_2}\varphi_{II}(\theta{-}\theta')\log(1+e^{-f(\theta')})\,d\theta'
\label{repnlie}
\end{eqnarray}
where
\begin{equation}
\varphi_{II}(\theta)=\varphi(\theta)+\varphi(\theta{-}i\pi)=
\frac{M\sin(\pi M)}{\pi(\cosh(2M\theta{-}i\pi M)-\cos(\pi M))}~.
\end{equation}
This holds for $\pi< \Im m\,\theta<\pi/M$.

The goal is to locate
the zeroes of $T(E)$ on or near the negative-$E$ axis. To this end
we set $\theta=i\pi(M{+}1)/(2M)+\gamma$, define the shifted counting
function
\begin{equation}
g(\gamma):=f(i\pi(M{+}1)/(2M)+\gamma)
\label{gdef}
\end{equation}
Abusing the notation a little, we shall sometimes write $g$ as a function
of the energy $E$, $g(E):=f(-E)$, where
\begin{equation}
E=
\left(re^{\gamma}\right)^{2M/(M{+}1)}.
\label{evalloc}
\end{equation}
The eigenvalues of the $\PT$-symmetric problem (\ref{ush}) are then
those values of $E$ such that
\eq
g(E)=(2k{+}1)\pi i
\label{econd}
\en
for some $k\in\Z$\,.

Defining $\psi(\gamma):=\varphi_{II}(i\pi(M{+}1)/(2M)+\gamma)$,
$g(\gamma)$ is given exactly,
for $|\Im m\,\gamma|<\frac{\pi}{2}|\frac{M{-}1}{M}|$\,,
by the following expressions:\\

\noindent $M>1$:
\begin{eqnarray}
g(\gamma)&=& 2i \sin(\fract{\pi}{2M})\,mre^{\gamma}
+\int_{C_1}\psi(\gamma{-}\theta')\log(1+e^{f(\theta')})\,d\theta'
\nn\\[3pt]
&&\qquad\qquad\qquad\qquad\qquad -
\int_{C_2}\psi(\gamma{-}\theta')\log(1+e^{-f(\theta')})\,d\theta'
\qquad
\label{gattrnlie}
\end{eqnarray}
with
\begin{equation}
\psi(\gamma)=
\frac{2\,\cos(\frac{\pi}{2M})\cosh(\gamma)}%
{\pi(\cosh(2\gamma)+\cos(\frac{\pi}{M}))}
=
\sum_{n=1}^{\infty}
(-1)^{n+1}\frac{2}{\pi}\,\cos\bigl(\fract{\pi}{2M}(2n{-}1)\bigr)\,
e^{-(2n{-}1)\gamma}\,.
\label{psiattr}
\end{equation}

\noindent
$M<1$:
\begin{eqnarray}
g(\gamma)&=&2i\pi(l{+}\fract{1}{2})
+\int_{C_1}\psi(\gamma{-}\theta')\log(1+e^{f(\theta')})\,d\theta'
\nn\\[3pt]
&&\qquad\qquad\qquad\qquad-
\int_{C_2}\psi(\gamma{-}\theta')\log(1+e^{-f(\theta')})\,d\theta'
\label{grepnlie}
\end{eqnarray}
with
\begin{equation}
\psi(\gamma)=
\frac{-M\sin(\pi M)}{\pi(\cosh(2M\gamma)+\cos(\pi M))}
=\sum_{n=1}^{\infty}
(-1)^n\frac{2M}{\pi}\,\sin\bigl(\pi Mn\bigr)\,
e^{-2Mn\gamma}\,.
\label{psirep}
\end{equation}
(We have recorded the large-$\gamma$ expansions for
$\psi$ as they will be needed later.)

The formulae (\ref{gattrnlie}) and (\ref{grepnlie})
express $g(\gamma)$ in terms of the values of
$f(\theta')$ on the contours $C_1$ and $C_2$, where the original
(`first determination') specification (\ref{nlie}) applies.
For a given value of $\gamma$, the dominant contributions
to the integrals in
(\ref{gattrnlie}) and (\ref{grepnlie}) come when $\Re
e\,\theta'\approx\Re e\,\gamma$, since the kernel $\psi(\gamma)$
is peaked about $\gamma=0$.
For large $\Re e\,\theta'$ on $C_1$ and $C_2$, $f(\theta')\sim
i\pi(l{+}1)-imre^{\theta'}$. Since $C_1$ is just below the real axis,
and $C_2$ just above, in this limit
$\log(1+e^{f(\theta')})\to 0$ on $C_1$, and
$\log(1+e^{-f(\theta')})\to 0$ on $C_2$.
Therefore, for large $\Re e\,\gamma$ the integrals in
(\ref{gattrnlie}) and (\ref{grepnlie}) can be dropped,
giving the following leading 
approximations\footnote{The result for $M=1$,
which is exact, follows directly from (\ref{nlie})
since $\varphi(\theta)\equiv 0$ in this case and no second
determination is required.  Since $m|\php_{M=1}=\pi/2$, it 
interpolates between the results for $M=1^+$ and $M=1^-$. This
interpolation will be discussed in more detail
around equation (\ref{smooth}) below.}
 for $g(\gamma)$\,:
\begin{eqnarray}
M>1~:~ && g(\gamma)\sim 2i \sin(\fract{\pi}{2M})m
E^{\frac{(M{+}1)}{2M}}
\label{attrlead}\\[3pt]
M=1~:~ && g(\gamma) = i\pi(l{+}\fract{1}{2})+ i\fract{\pi}{2}E
\label{fflead}\\[3pt]
M<1~:~ && g(\gamma)\sim 2i\pi(l{+}\fract{1}{2})
\label{replead}
\end{eqnarray}

For $M>1$, the leading approximation for the energy levels is
obtained by setting
the RHS of (\ref{attrlead}) equal to
$(2n{+}1)\pi i$\,.  Then, using (\ref{mdef}),
the WKB result of \cite{Bender:1998ke,Bender:1998gh} is recovered:
\begin{equation}
E_n\sim
\,\left(\frac{\sqrt{\pi}\,\Gamma\left(\frac{3}{2}{+}\frac{1}{2M}\right)\,%
(n{+}\fract{1}{2})}{\sin(\fract{\pi}{2M})\,%
\Gamma\left(1{+}\fract{1}{2M}\right)}\right)^{2M/(M{+}1)}.
\label{wkbres}
\end{equation}

For $M=1$, there are some subtleties. Since the eigenvalues at this
point are easily found exactly, they can be ignored, but their
resolution is perhaps interesting. At $M=1$, $\omega=i$ and the
factors $Q(\omega^{\pm 2})$ in (\ref{adef}) cancel, making $f$ and
$g$ as naively defined by (\ref{fdef}) and (\ref{gdef}) equal to
$\pi(l+\frac{1}{2})$. However this is too quick: at $M=1$ the TQ
relation (\ref{tq}) is `renormalised' to
$T(E)\,Q(E)=\omega^{-l-1/2+E/2}\,Q(\omega^{-2}E)+
\omega^{l+1/2-E/2}\,Q(\omega^{2}E)$
\cite{Bazhanov:1996dr,Dorey:1999uk}, and
as a result (\ref{adef}) is naturally replaced by
$a(E)=\omega^{2l+1-E}Q(\omega^2E)/Q(\omega^{-2}E)$, which indeed
matches (\ref{fflead}). Next, the renormalised TQ relation
implies $T(E)Q(E)=2\cos((2l{+}1{-}E)\pi/4)Q(-E)$.
Knowing that the zeroes of $Q$ are positive and those of $T$ negative
allows this relation
to be disentangled: the zeroes of $Q$ must be at
$E=2l-1+4n$, and those of $T$ at
$E=2l+3-4n$ and $-2l+1-4n$, for $n=1,2,\dots$\,. Recalling that the
eigenvalues of the lateral problem are at the negated zeroes of $T$
recovers the spectrum of the `PT-symmetric simple harmonic oscillator',
previously obtained in \cite{Znojil:1999qt} and
\cite{Dorey:1999uk} via explicit solutions of the differential
equation.

Finally, for $M<1$
the leading approximation to $g(\gamma)$ is a constant, and does not
give any information about the energy levels at all.
This gives a novel insight into the observation of~\cite{Bender:1998ke}
that standard WKB techniques fail for $M<1$\,: from the point of view
of the ODE/IM correspondence and its associated nonlinear integral
equation, it can be traced to the change in the nature of the second
determination in the repulsive regime.

\[
\!\!\!
\begin{array}{ccc}
\includegraphics[width=0.29\linewidth]{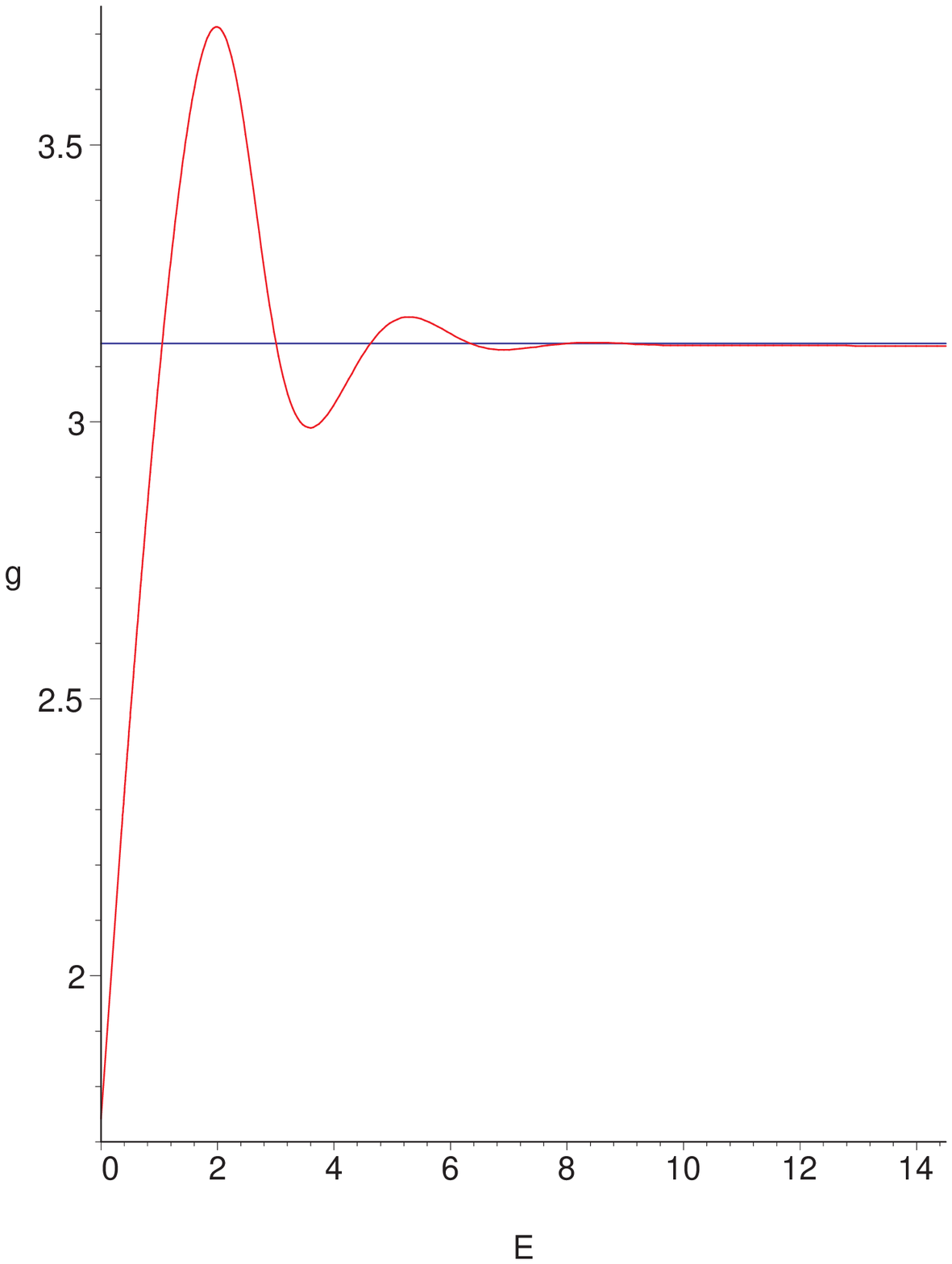}~~&~~%
\includegraphics[width=0.29\linewidth]{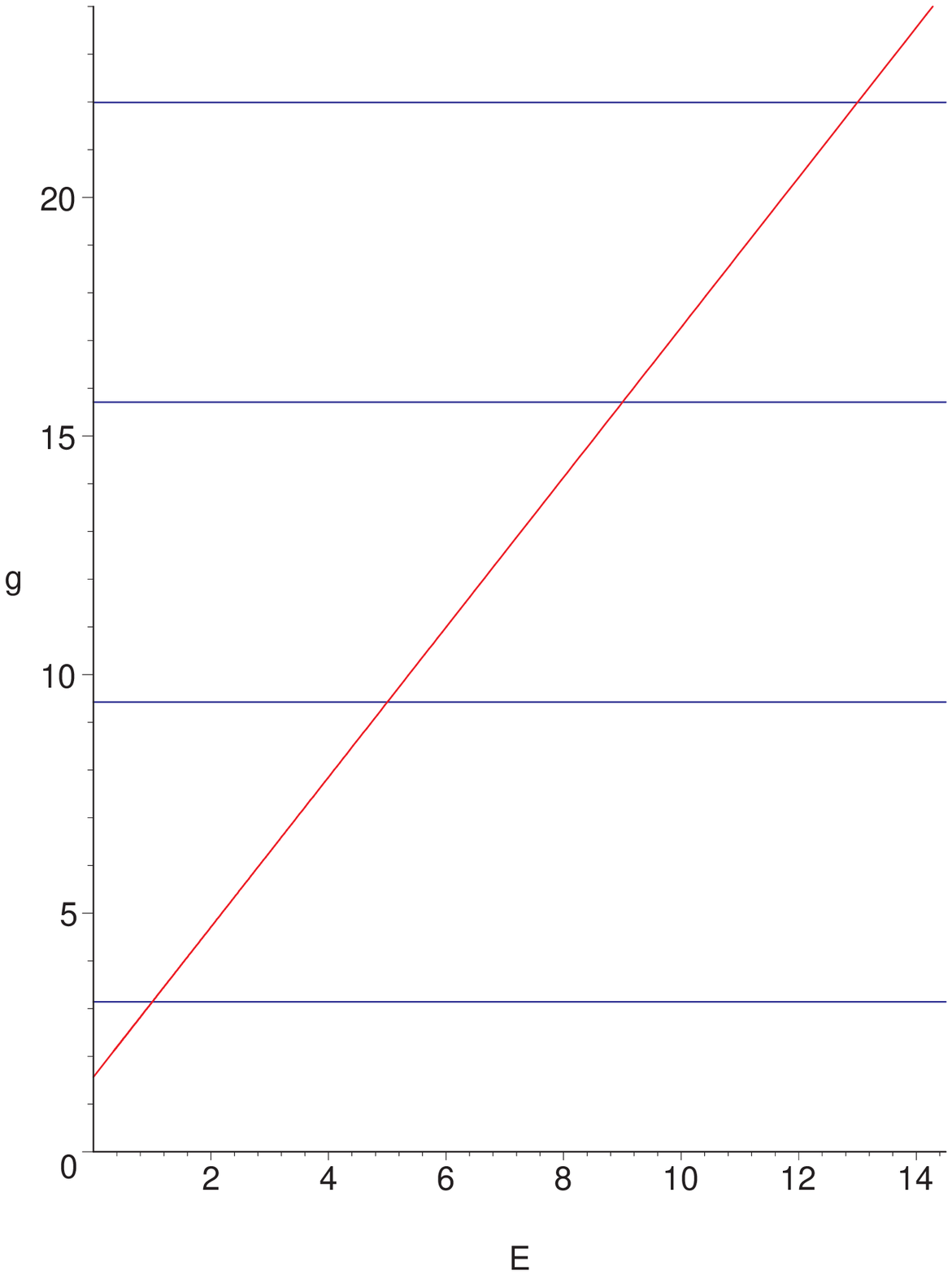}~~&~~%
\includegraphics[width=0.29\linewidth]{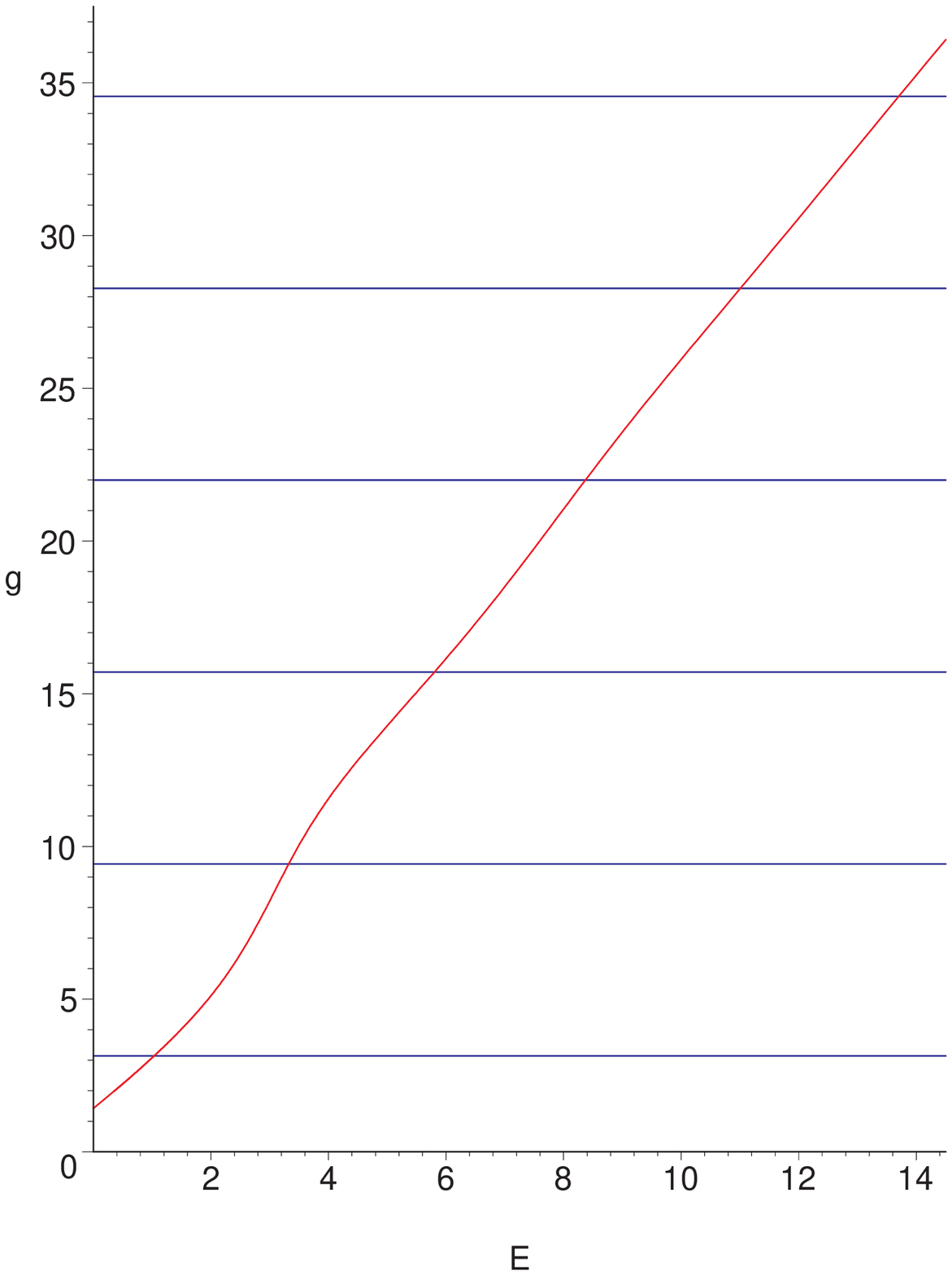}
\\[4pt]
\parbox{0.24\linewidth}{~~~~~\small \ref{seconddets}a:
$M=0.8$}&
\parbox{0.24\linewidth}{~~~~~~\small \ref{seconddets}b:
$M=1$}&
\parbox{0.24\linewidth}{~~~~~~~\small \ref{seconddets}c:
$M=1.2$}\\[9pt]
\multicolumn{3}{c}%
{
\parbox{0.92\linewidth}{
Figure \protect\ref{seconddets}: 
$\Im m\,g(E)$ from the nonlinear integral equation.
In all three cases, $l=-0.001$. The horizontal lines show the
`quantisation levels' $g=(2n{+}1)\pi i$.
}
}
\end{array}
\refstepcounter{figure}\label{seconddets}
\]

To proceed further, the more detailed behaviour of $g(E)$ is required. 
Figure \ref{seconddets} shows $\Im m\,g(E)$ in the
various regimes, the plots being obtained by solving the NLIE 
numerically, but with no other approximations. Note that the
nature of the energy level quantisation is radically changed 
for $M<1$: even for the levels which are still real, there is
no natural way to allot them a unique WKB (Bohr-Sommerfeld) quantum 
number via the counting function, as happens for
standard eigenvalue problems in quantum mechanics, and also for 
these lateral ($\PT$-symmetric) problems when $M>1$.

The next task is to understand figure
\ref{seconddets} analytically.
It turns out that the direct use of the asymptotic
expansion for $Q(E)$ developed in \cite{Bazhanov:1996dr,Bazhanov:1998za}
fails to capture most of the important
features of the energy levels for $M<1$\,: roughly put, the wiggles in
figure \ref{seconddets}a are missed, and the asymptotic behaviour of just
a single level is captured -- the partner of the ground
state, which moves up to infinity as $l\to 0^-$ (this can be seen on
figure \ref{numerics}b, and is illustrated
in more detail in figure \ref{pertcomp}a below).
The remaining levels only emerge when a `beyond all
orders' term is added. The need for this term is
most easily understood if we continue to work directly with the
nonlinear integral equation.

Higher corrections to $g(\gamma)$ can be obtained through a
steepest-descent treatment of the integrals in
(\ref{gattrnlie}) or (\ref{grepnlie}), expanding the kernels
$\psi(\gamma-\theta')$ for large $\gamma-\theta'$ and using again the
fact that the terms $\log(1+e^{f(\theta')})$ and
$\log(1+e^{-f(\theta')})$ decay to zero as $\Re e\,\theta'\to\infty$
along $C_1$ and $C_2$ respectively. This requires
the contours $C_1$ and $C_2$ to be shifted so as to
maximise the rates of decay of $\log(1+e^{\pm f(\theta')})$ along
them. For large $\Re e\,\theta'$ with $|\Im
m\,\theta'|<\min(\pi,\pi/M)$\,, $f(\theta')\sim
i\pi(l{+}\fract{1}{2})-imre^{\theta'}$, so
$C_1$ should be shifted down to $\Im m\,\theta'=-\pi/2$\,, and
$C_2$ shifted up to $\Im m\,\theta'=+\pi/2$\,. (For $M<2$, these
lines do indeed lie inside the first determination, for which the
just-mentioned asymptotic of $f(\theta')$ holds, and  we restrict to
such cases in the following.)
These shifts of contours are not entirely innocent operations, as the
kernel functions $\psi(\gamma)$ have poles at
$\gamma=\pm\frac{i\pi}{2}\frac{|M{-}1|}{M}$\,. Taking $C_1$ and $C_2$
past these poles generates residue terms proportional to the values of
$\log(1+e^{\pm f(\theta')})$ there. These are the `beyond all orders'
terms just advertised.

Explicitly, let $\widetilde C_1$ and $\widetilde C_2$ be the contours $C_1$
and $C_2$ shifted onto the steepest descent (for large $\Re
e\,\theta'$) paths $\Im m\,\theta'=-\pi/2$ and
$\Im m\,\theta'=+\pi/2$ respectively. For $1<M<2$\,,
(\ref{gattrnlie}) can then be rewritten as
\begin{eqnarray}
g(\gamma)&\!=\!& 2i\sin(\fract{\pi}{2M})\,mre^{\gamma}\nn\\[4pt]
&&{}~~
+\int_{\widetilde
C_1}\psi(\gamma{-}\theta')\log(1+e^{f(\theta')})\,d\theta'
 -\int_{\widetilde
C_2}\psi(\gamma{-}\theta')\log(1+e^{-f(\theta')})\,d\theta'
\nn\\[4pt]
&&{}~~~~
 +\log(1+e^{f(\gamma-\frac{\pi i}{2}\frac{|M-1|}{M})})
 -\log(1+e^{-f(\gamma+\frac{\pi i}{2}\frac{|M-1|}{M})})\,.
\end{eqnarray}
For $\frac{1}{2}<M<1$, the representation
(\ref{grepnlie}) is rewritten in a similar manner,
modulo a change in the signs of the extra terms, due to the opposite
residues of the relevant poles.
(For $M<\frac{1}{2}$, the poles are not encountered in the
shifting of $C_1$ and $C_2$, and so no extra terms are needed.)

Once the contours have been shifted, the kernel functions are
expanded using (\ref{psiattr}) and (\ref{psirep})
to give asymptotic expansions of the integrals in the
standard way. 
Setting $E=(re^{\gamma})^{2M/(M{+}1)}$,
for $1<M<2$ this yields
\begin{eqnarray}
\!\!\!\!
\!\!\!\!
g(E)&\!\sim\!& 2i\sin(\fract{\pi}{2M})m
E^{\frac{(M{+}1)}{2M}}\nn\\[3pt]
&&~+i\sum_{n=1}^{\infty}
(-1)^{n+1}\,\frac{2\,b_n}{\pi}\,\cos\bigl(\fract{\pi}{2M}(2n{-}1)\bigr)\,
E^{-\frac{(M{+}1)}{2M}(2n{-}1)}
{}-g\php_{\rm nonpert}(E)\quad
\label{attrexp}
\end{eqnarray}
where 
the coefficients $b_n$ are given in terms of
$f(\theta)$ by
\begin{equation}
ib_n=
\int_{C_1}(re^{\theta})^{2n{-}1}\log(1+e^{f(\theta)})\,d\theta
 -\int_{C_2}(re^{\theta})^{2n{-}1}\log(1+e^{-f(\theta)})\,d\theta
\label{bcoeffs}
\end{equation}
and
\eq
g\php_{\rm nonpert}(E)=
\log\left(\frac%
{1+e^{-f(\gamma+\frac{\pi i}{2}\frac{|M-1|}{M})}}%
{1+e^{f(\gamma-\frac{\pi i}{2}\frac{|M-1|}{M})}}%
\right)\,.
\label{gnp}
\en
(After the expansion has been made, the contours in (\ref{bcoeffs})
can be
shifted back to their original locations without encountering any
poles.) The geometrical interpretation of $g\php_{\rm nonpert}$ will
be clearest if the branch of the logarithm in (\ref{gnp}) is taken
in the interval $[-\pi,\pi)$, and this will be assumed from now on; 
note that this choice has no effect on
the quantization condition on the eigenvalues.

For $\frac{1}{2}<M<1$, the result is instead
\eq
g(E)\sim2i\pi(l{+}\fract{1}{2}) + i\sum_{n=1}^{\infty}
(-1)^n\,\frac{2Mc_n}{\pi}\,\sin\bigl(\pi Mn\bigr)\,
E^{-(M{+}1)n}+g\php_{\rm nonpert}(E)
\label{repexp}
\en
with coefficients $c_n$ given by
\begin{equation}
ic_n=
\int_{C_1}(re^{\theta})^{2Mn}\log(1+e^{f(\theta)})\,d\theta
 -\int_{C_2}(re^{\theta})^{2Mn}\log(1+e^{-f(\theta)})\,d\theta
\end{equation}
and
$g\php_{\rm nonpert}(E)$ is again given by (\ref{gnp}). Note
that $g\php_{\rm nonpert}$ 
now contributes with the opposite sign, while the modulus signs
in the formula mean that the numerator is always evaluated above the
real axis, and the denominator below, if $\gamma$ is real.

The integrals for the coefficients $b_n$ and $c_n$ are known
exactly in terms of the ground state eigenvalues of the
local and non-local conserved charges of the zero-mass
limit of the sine-Gordon model on a
circle~\cite{Bazhanov:1994ft,Bazhanov:1996dr}.
Translating\footnote{The dictionary
\cite{Dorey:1998pt,Bazhanov:1998wj} is $\beta^2=1/(M{+}1)$,
$p=(2l{+}1)/(4M{+}4)$\,.}
the results from
these papers into the current normalisations,
\begin{eqnarray}
b_n&=&\frac{\pi^{3/2}}{n!}\frac{(4M{+}4)^{n}}{2n{-}1}%
\,\frac{\Gamma((\frac{1}{2}{+}\frac{1}{2M})(2n{-}1))}%
{\Gamma(\frac{1}{2M}(2n{-}1))}%
\,I_{2n-1} ~;\quad\\[4pt]
c_n&=& -\frac{2\pi}{M}\,\frac{
2^{2Mn}}{(M{+}1)^{2n}}
\cos(\pi Mn)
\,\widetilde H_n~.
\end{eqnarray}
Here, $I_{2n-1}$
and $\widetilde H_n$ are respectively the ground state eigenvalues
of ${\mathbb I}_{2n-1}$\,,
the $n^{\rm th}$ local conserved charge,
and $\widetilde {\mathbb H}_n$\,, the $n^{\rm th}$ (dual)
nonlocal conserved charge. The precise definitions of these charges
can be found in \cite{Bazhanov:1994ft,Bazhanov:1996dr,Bazhanov:1998za} 
(see also \cite{Fioravanti:2002sq,Feverati:2004bv}).
Explicit expressions for the local charge
eigenvalues can be found up to high order; the first three are
\begin{eqnarray}
I_1&=&
\frac{1}{(4M{+}4)}%
\bigg[\lambda^2-\frac{M{+}1}{6}\bigg]~,\nn\\[4pt]
I_3&=&\frac{1}{(4M{+}4)^2}
\bigg[\lambda^4-(M{+}1)\lambda^2-\frac{(M{+}1)(4M{+}3)(M{-}3)}{60}%
\bigg]~,\nn\\[6pt]
I_5&=& \frac{1}{(4M{+}4)^3}\bigg[\lambda^6
-\frac{5(M{+}1)}{2}\lambda^4-
\frac{(M{+}1)(4M^2{-}23M{-}23)}{12}\lambda^2\nn\\[3pt]
&&\qquad\qquad\qquad
{}-\frac{(M{+}1)(96M^4{-}340M^3{+}85M^2{+}850M{+}425)}{1512}\bigg]~,
\label{loccharges}
\end{eqnarray}
where $\lambda=l{+}\frac{1}{2}$\,.
The nonlocal charge eigenvalues are harder to calculate, but the
first one is also known in closed form:
\begin{equation}
\widetilde H_1=
\frac{(M{+}1)^2\,\Gamma(M{+}1)\Gamma(-2M{-}1)\Gamma(M{+}1{+}\lambda)}%
{\Gamma(-M)\Gamma(-M{+}\lambda)}~.
\label{nlval}
\end{equation}

Finally, the beyond-all-orders term 
$g\php_{\rm nonpert}(E)$ in 
(\ref{attrexp}) and (\ref{repexp}) should be treated, 
which contains the
so-far unknown function $f(\theta)$. 
We first note that for $\gamma$ real
$f(\gamma-\frac{\pi i}{2}\frac{|M-1|}{M})=
-f^*(\gamma+\frac{\pi i}{2}\frac{|M-1|}{M})$ (the ${}^*$ denoting
complex conjugation) and so
\eq
g\php_{\rm nonpert}(E)=2\,i\arg\left(1+e^{-f(\gamma+\frac{\pi
i}{2}\frac{|M-1|}{M})}\right).
\label{gnpb}
\en
Given the branch chosen for the logarithm just after (\ref{gnp}), the
value of the argument in (\ref{gnpb}) lies between $-\pi$ and $\pi$.
In principle the full asymptotic 
expansion for $f$, to be given in the next section, could now be
substituted into this formula. However the interesting 
structure is already seen if the leading behaviour is used, which,
from (\ref{nlie}), is
\eq
f(\gamma+\fract{\pi i}{2}\fract{|M-1|}{M}))
\sim i\pi(l{+}\fract{1}{2})-im\,e^{\frac{\pi i}{2}\frac{|M-1|}{M}}%
E^{\frac{M+1}{2M}}.
\en
Hence
\eq
g\php_{\rm nonpert}(E)\sim
2\,i\arg\left(1+
\rho\, e^{i\phi}
\right)
\label{gnpt}
\en
where
\bea
\rho(E)&=&
e^{-m\sin\!\frac{\pi|M-1|}{2M}\, E^{\frac{M+1}{2M}}}\,;
\\[6pt]
\phi(E)
&=&-\pi(l{+}\fract{1}{2})+m\cos\!\fract{\pi|M{-}1|}{2M}\,E^{\frac{M+1}{2M}}\,.
\eea
The final formula can be given a pictorial interpretation:
$g_{\rm nonpert}(E)$ is approximately equal to $2i\,\chi(E)$\,,
where $\chi(E)$ the interior angle of the triangle
shown in figure \ref{trigfig}.

\[
\begin{array}{c}
\begin{picture}(200,77)(-15,-20)
\thinlines
\put(20,0){\vector(1,0){100}}
\put(120,0){\line(1,0){15}}
\put(120,0){\vector(-4,3){64}}
\put(56,48){\line(-3,-4){36}}
\put(65,-6){\makebox(0,0)[t]{$1$}}
\put(87,31){\makebox(0,0)[bl]{$\rho(E)$}}
\put(120,4){\makebox(0,0)[bl]{$\phi(E)$}}
\put(34,4){\makebox(0,0)[bl]{$\chi(E)$}}
\end{picture}\\
\parbox{0.45\linewidth}{\small
Figure \ref{trigfig}: The trigonometry of $g_{\rm nonpert}(E)$\,.}
\end{array}
\refstepcounter{figure}\label{trigfig}
\]

Notice that $\rho(E)$ is always less than $1$ for $M\neq 1$, and that as
$E$ increases with $M$ fixed
the point $1+\rho\,e^{i\phi}$ executes a diminishing
spiral about $1$, giving $\chi(E)$ a damped oscillation between
$-\pi/2$ and $\pi/2$ which captures the behaviour seen for $M<1$
in figure \ref{seconddets}a above.
The nonperturbative nature of $g_{\rm nonpert}(E)$ is 
easily seen: $|\chi(E)|$ is bounded by $\frac{\pi}{2}\rho(E)\,$;
and at fixed $M$, the expansion
parameters for the perturbative series parts of
(\ref{attrexp}) and
(\ref{repexp}) are $\varepsilon:=E^{-\frac{M{+}1}{2M}}$ and
$\eta:=E^{-(M{+}1)}$ respectively, in terms of which $\frac{\pi}{2}\rho$ 
is equal
to either $\frac{\pi}{2}\exp(-m\sin\!\frac{|M-1|}{M}/\varepsilon)$
or $\frac{\pi}{2}\exp(-m\sin\!\frac{|M-1|}{M}/\eta^{\frac{1}{2M}})$\,.

If on the other hand $M$ is taken very close to $1$ with $E$ remaining
finite, the nonperturbative term plays a key role. In this
limit, $\rho(E)\approx 1$ and hence, from figure
\ref{trigfig}, $\chi(E)\approx\phi(E)/2$,
and
\eq
g\php_{\rm nonpert}(E)\approx
-i\pi(l{+}\fract{1}{2})+im\,E^{\frac{M+1}{2M}}\,.
\label{smooth}
\en
This is just the behaviour required if the
nonperturbative term is to `smooth' the discontinuous change in the
leading asymptotic of $g(E)$ as $M$ passes through $1$, seen in 
equations (\ref{attrlead}) --
(\ref{replead}) above. The phenomenon is reminiscent of the smoothing
of Stokes's discontinuities discussed in \cite{Berry}.
Strictly speaking, since 
$\chi$ remains between
$-\pi/2$ and $\pi/2$ the limit of 
$g\php_{\rm nonpert}(E)/i$ as $M\to 1$
is a sawtooth function obtained by returning
the imaginary part of the RHS of (\ref{smooth}),
modulo $2\pi$, to the interval $[-\pi,\pi]$.
This has an important effect on the counting of eigenvalues. Consider
for example the limit
$M\to 1^-$ at fixed $E$: since
the coefficients in series part of (\ref{repexp}) are all of order
$\epsilon=2-2M$, the limiting form of $g(E)$ is
\eq
g(E)\Big|\php_{M=1^-} = 2\pi i(l+\fract{1}{2})-\pi i
\left[\,l+\fract{1}{2}-\fract{1}{2}E\,\right]_{[-1,1]}
\en
where
the subscript on the last term indicates that it should be returned,
modulo $2$,
to the interval $[-1,1]$ -- so $[x]_{[-1,1]}:=(x{+}1)\,{\rm mod}\,2 -1$. 
Since $g$ is continuous
for all $M<1$, its $M\to 1^-$ limit has
segments of infinite gradient at the points
$l+\frac{1}{2}-\frac{1}{2}E=(2k{+}1)$
where the corresponding sawtooth
function would have a discontinuity. The quantisation condition
$g(E)=(2n{+}1)\pi i$ is therefore {\em always}\/ satisfied at these
values of $E$, in addition to the points
$l+\frac{1}{2}+\frac{1}{2}E=(2k{+}1)$ which would have been found even
before taking the sawtooth effect into account. Taken together, these
points match the
full spectrum of the PT-symmetric
simple harmonic oscillator, discussed from
another point of view just after equation (\ref{wkbres}) above;
the crucial role played by the nonperturbative term in this
process is interesting.

\[
\!\!\!
\begin{array}{ccc}
\includegraphics[width=0.29\linewidth,height=0.5\linewidth]%
{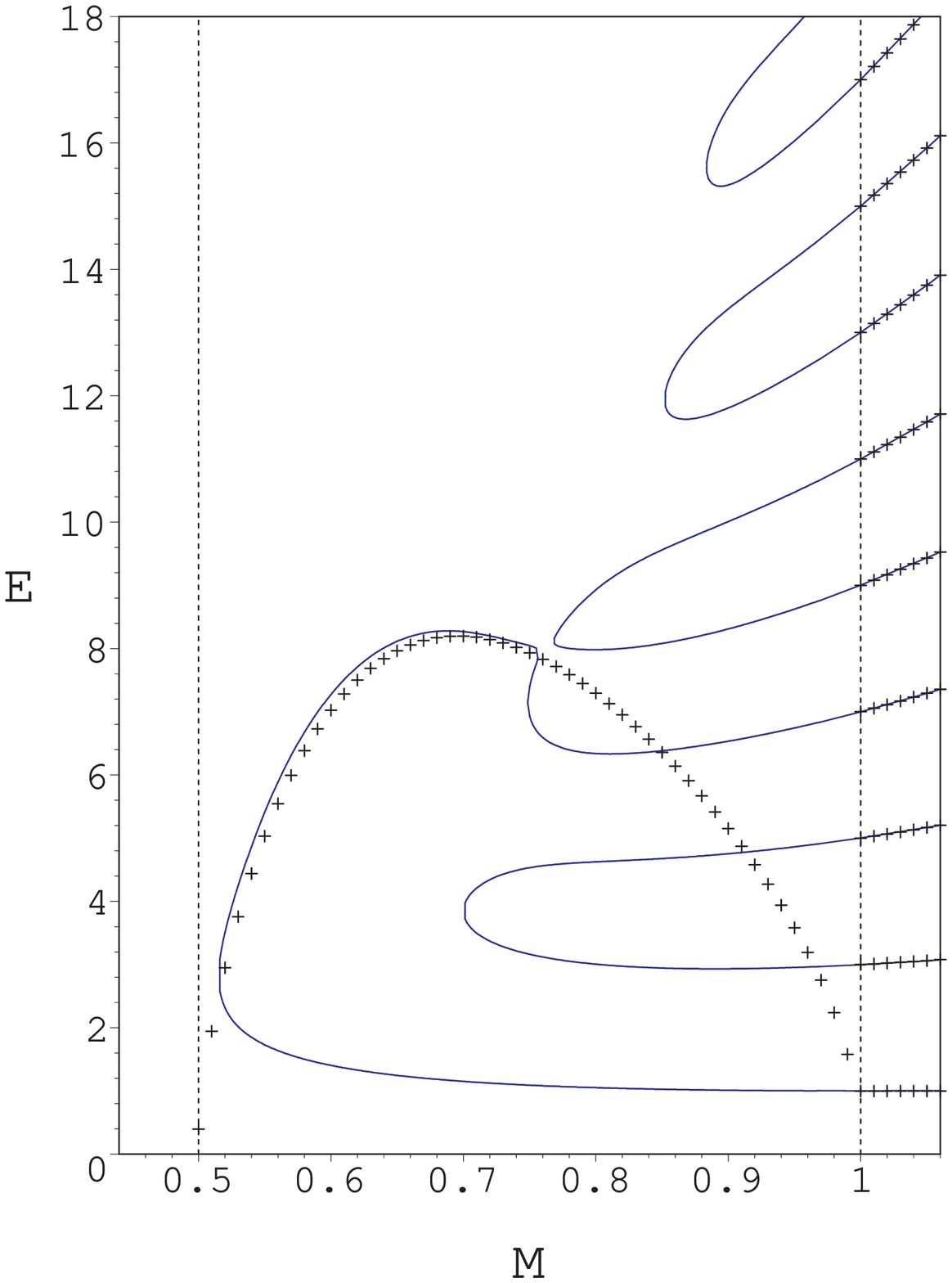}~~&~~%
\includegraphics[width=0.29\linewidth,height=0.5\linewidth]%
{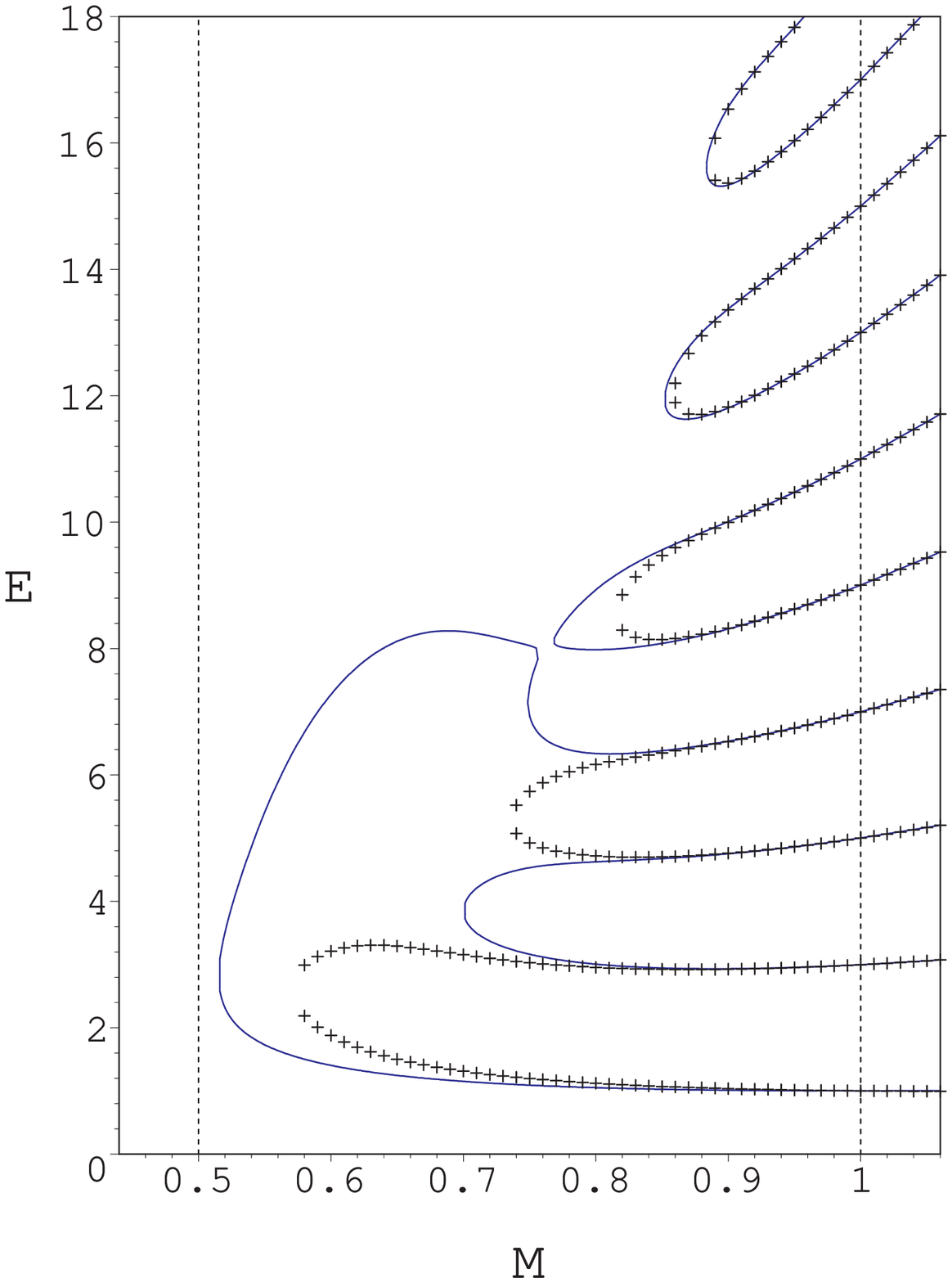}~~&~~%
\includegraphics[width=0.29\linewidth,height=0.5\linewidth]%
{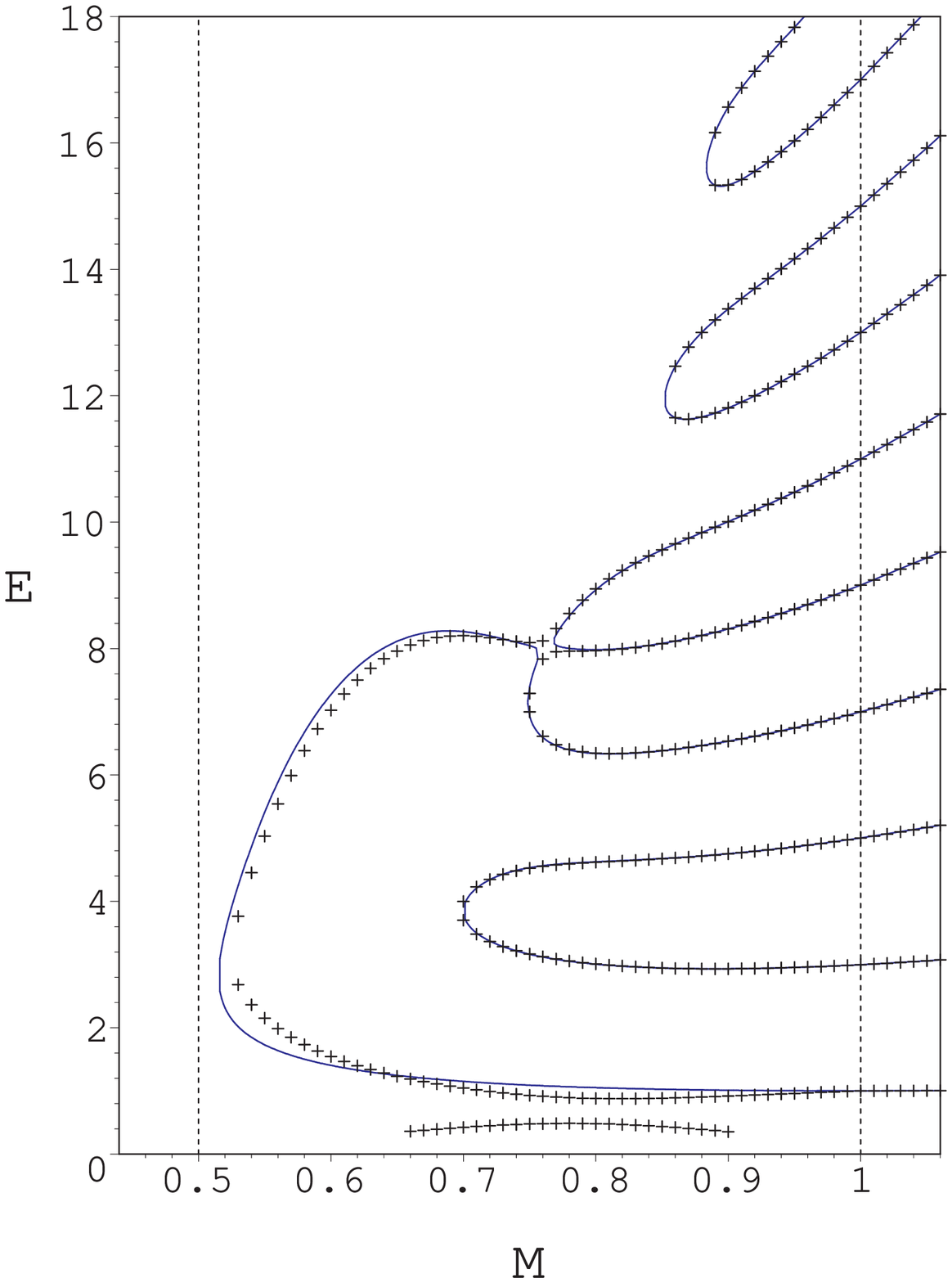}
\\[4pt]
\parbox{0.24\linewidth}{\small \protect\ref{pertcomp}a:
Perturbative term}&~~~
\parbox{0.28\linewidth}{\small \protect\ref{pertcomp}b:
Nonperturbative term}&~~
\parbox{0.24\linewidth}{\small \protect\ref{pertcomp}c:
Both terms}\\[9pt]
\multicolumn{3}{c}%
{
\parbox{0.9\linewidth}{\small
Figure \protect\ref{pertcomp}: 
Different approximation schemes for the energy levels 
(small crosses) compared with the exact levels (continuous lines), for
$l=-0.001$.
}
}
\end{array}
\refstepcounter{figure}\label{pertcomp}
\]

More generally, for $M<1$ and $l\neq 0$ the nonperturbative term explains
the infinite sequence of merging levels high in the spectrum,
while perturbative term is responsible for the single
level joined to the ground state which moves upwards as $l\to
0$. To understand the reversed connectivity below this level, both
effects must be incorporated simultaneously. 
Figure~\ref{pertcomp} illustrates this by comparing
the exact energy levels,
computed using the full nonlinear integral equation, 
with  the
levels obtained by combining the leading asymptotic of $g(E)$,
(\ref{attrlead}) - (\ref{replead}), with either a) the first
perturbative correction, or b) the leading approximation to the
nonperturbative correction, or c) both. (Note that 
the lowest group of crosses in
figure \ref{pertcomp}c at intermediate values of $M$ is spurious,
being caused by the over-large contribution of the perturbative
correction at such small values of $E$.)

Explicitly, the crosses on figure \ref{pertcomp}c 
mark points at which
$g_{\rm approx}(E)=(2k{+}1)\pi i$, where 
\eq
g_{\rm approx}(E)=
\left\{
\begin{array}{ll}
2im\sin(\fract{\pi}{2M})\,
E^{\frac{(M{+}1)}{2M}}
+
\frac{i\,\pi^{3/2}\,
(6(l{+}\frac{1}{2})^2-M{-}1)
}{3\,\Gamma(\frac{1}{2M})\Gamma(\frac{1}{2}-\frac{1}{2M})}%
\, E^{-\frac{(M{+}1)}{2M}}
{}-g\php_{\rm nonpert}(E)
&(M>1);\\[9pt]
2i\pi(l{+}\fract{1}{2})
+\frac{i\,\pi^{3/2}\,\Gamma(M{+}\frac{3}{2}{+}l)}%
{\Gamma(M{+}\frac{3}{2})\Gamma(-M)\Gamma(-M{+}\frac{1}{2}{+}l)}\,
E^{-(M{+}1)}+g\php_{\rm nonpert}(E)
&(M<1),
\end{array}
\right.
\label{econdapprox}
\en
with $g_{\rm nonpert}(E)$ given by the leading approximation
(\ref{gnpt}), and $m$ by (\ref{mdef}). 
For figure \ref{pertcomp}a, only the first and second
terms on each RHS were used for $g_{\rm approx}$; 
for figure \ref{pertcomp}b, only the first and 
third.

Even for the low-lying levels, the combined 
approximation does very well. 
Figure \ref{zooms} shows a more stringent test by
zooming in on the first three level-mergings; again, the match is very
good. 
The errors are greatest 
in figure \ref{zooms}b, near to the second merging.
This is not surprising
since the recombination of levels occuring there
makes the locations of the eigenvalues particularly sensitive to
errors in the value of $g(E)$. These errors become smaller for
recombinations higher up the spectrum.

\[
\!\!\!\!\!
\begin{array}{ccc}
\includegraphics[height=0.37\linewidth]%
{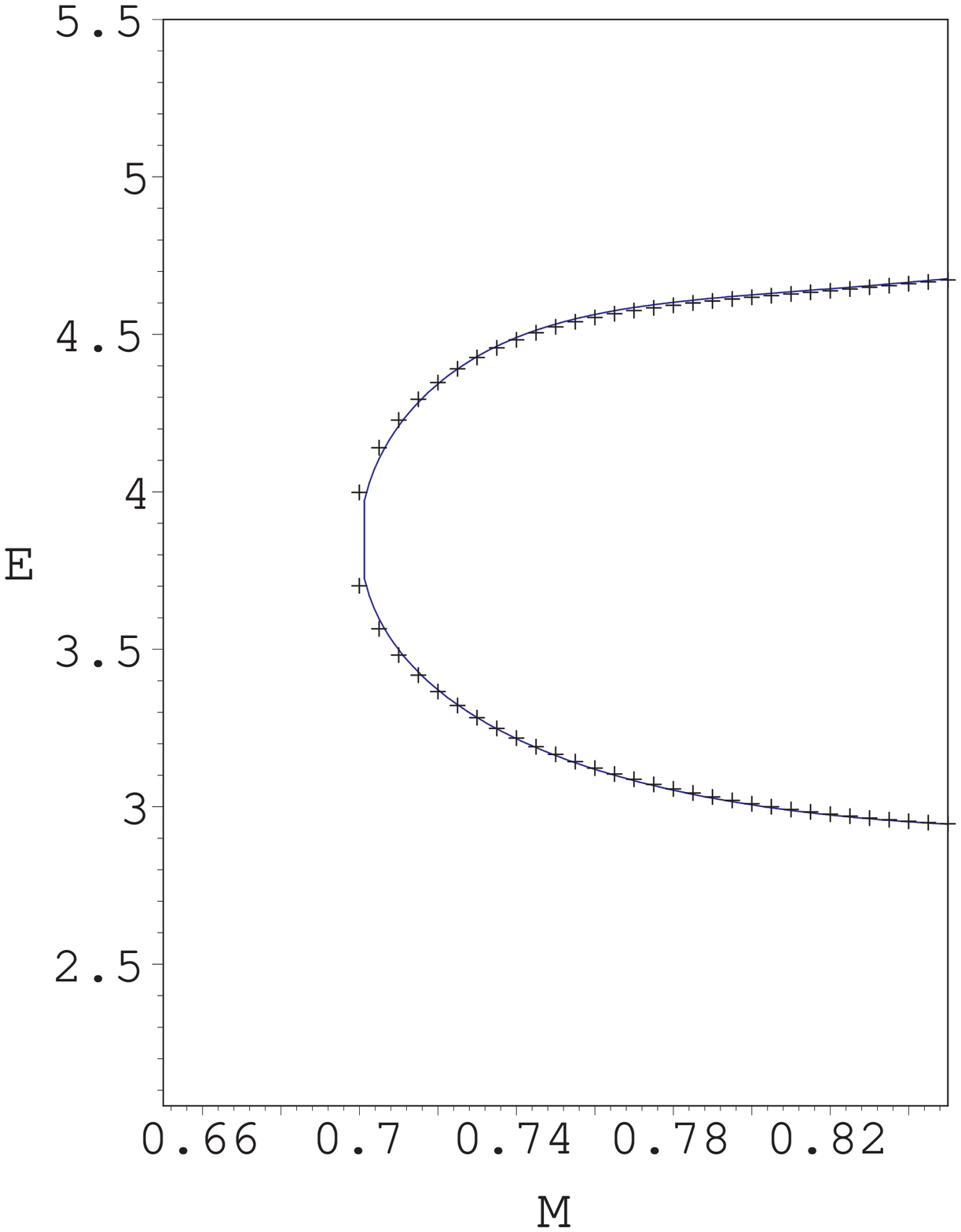}~~&~~%
\includegraphics[height=0.37\linewidth]%
{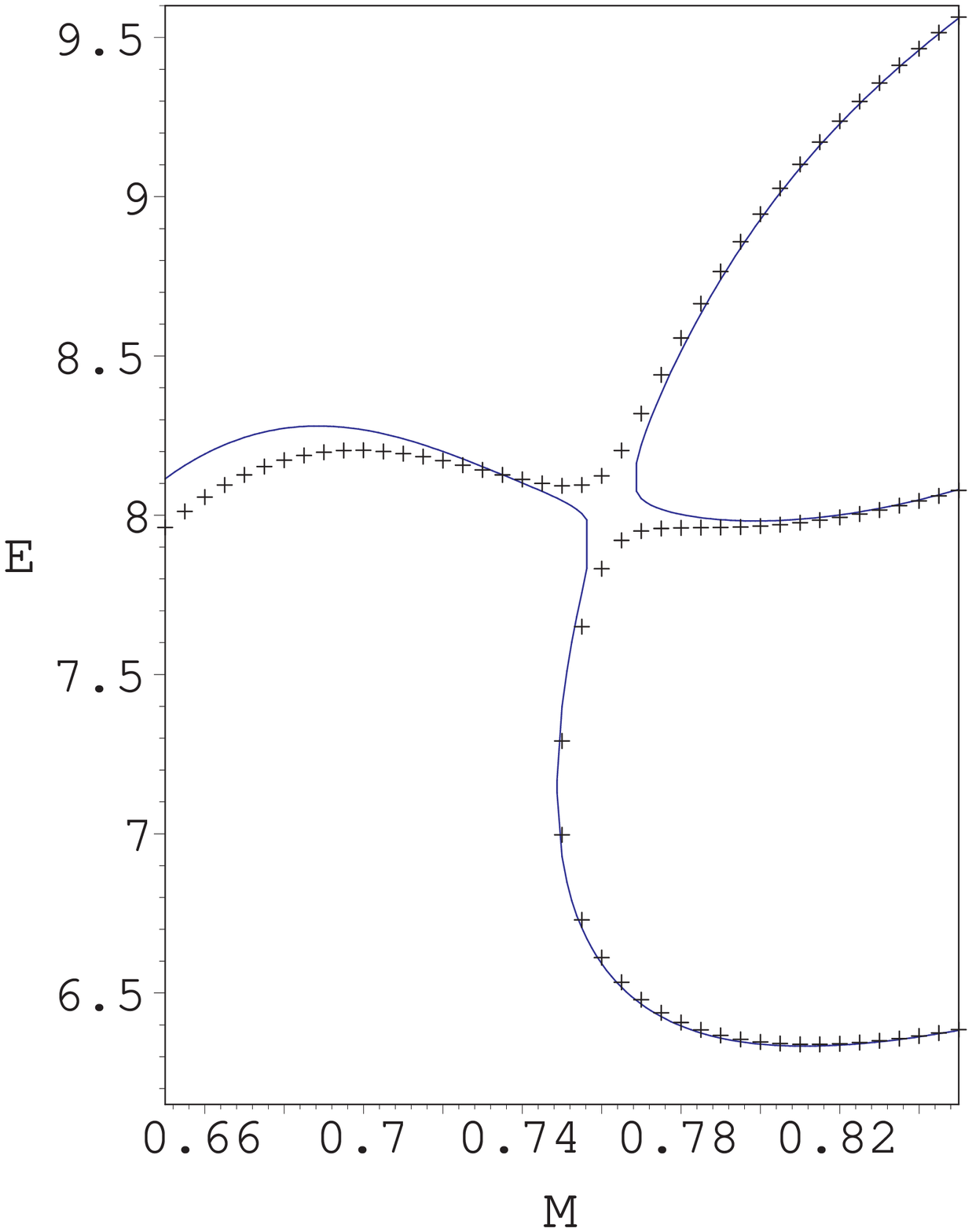}~~&~~%
\includegraphics[height=0.37\linewidth]%
{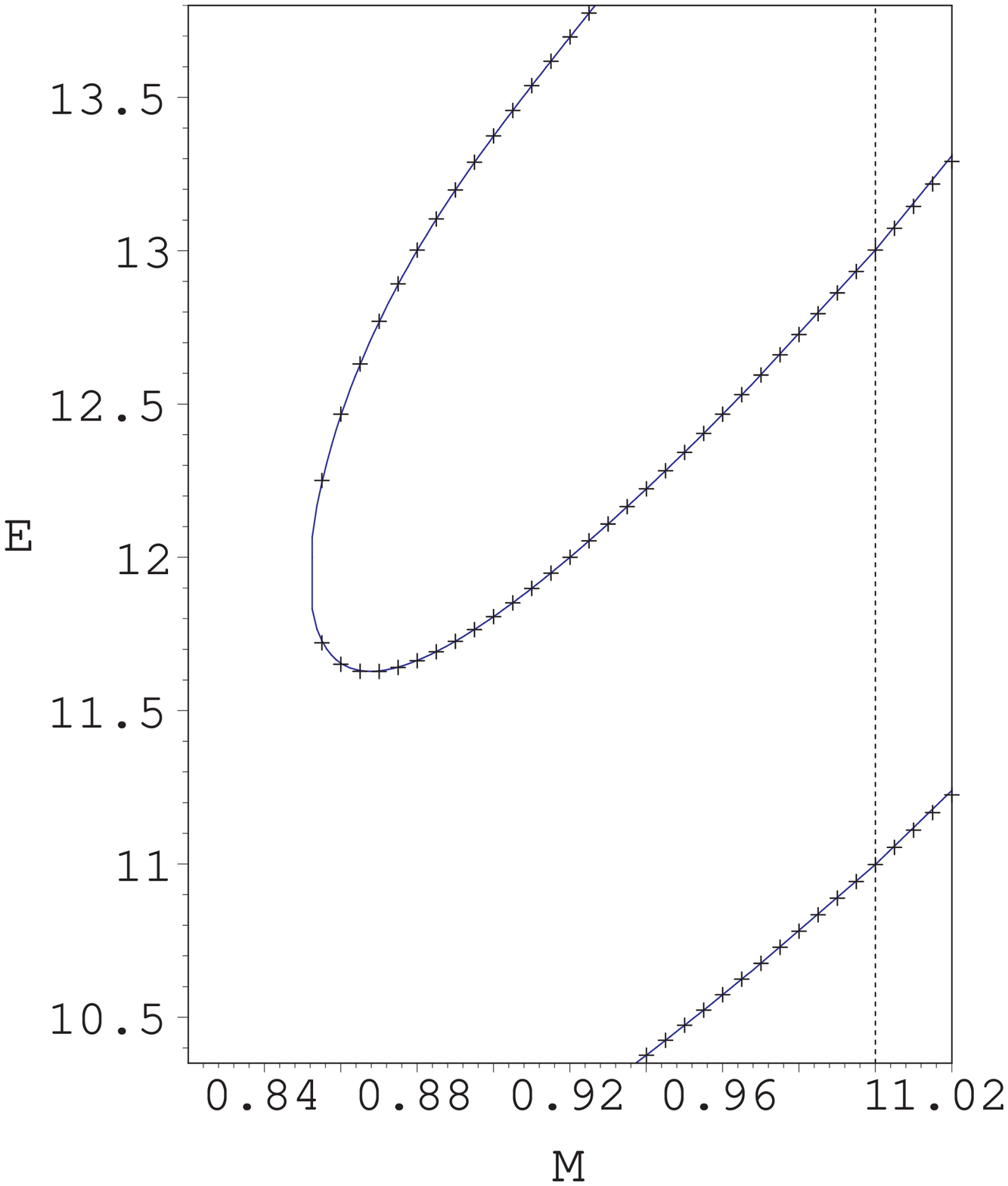}
\\[4pt]
\parbox{0.23\linewidth}{\small \protect\ref{zooms}a:
First merging}&~~
\parbox{0.23\linewidth}{\small \protect\ref{zooms}b:
Second merging}&~
\parbox{0.23\linewidth}{\small \protect\ref{zooms}c:
Third merging}\\[9pt]
\multicolumn{3}{c}%
{
\parbox{0.9\linewidth}{\small
Figure \protect\ref{zooms}: The
perturbative-plus-nonperturbative approximation
 for $l=-0.001$ (small crosses) 
compared with the exact levels (continuous lines) near the first
three level mergings 
shown on figure~\protect\ref{pertcomp}c.
}
}
\end{array}
\refstepcounter{figure}\label{zooms}
\]

To analyse the level-mergings for $M<1$ in more detail, we take
$\epsilon=2M{-}2$ small and negative, and $E$ fixed but large. Then
the limiting form of the quantisation condition arising from
(\ref{econdapprox}) can be written as
\eq
\chi(E)=\pi k-\pi l 
+\frac{\pi}{3}(l{+}\fract{3}{2})(l^2{-}\fract{1}{4})\,|\epsilon|\,E^{-2}
\label{econdlead}
\en
where $k\in\Z$\,, 
$\chi(E)\equiv\frac{1}{2i}\,g_{\rm nonpert}(E)$ is the 
angle shown on figure \ref{trigfig}, and, to the same approximation,
the other quantities on the figure are
\bea
\rho(E)&=&e^{-\frac{\pi^2}{8}|\epsilon|E}\,;
\label{rhoform}\\
\phi(E)&=&-\pi(l{+}\frac{1}{2})+\frac{\pi}{2}E+
\frac{\pi}{8}|\epsilon|\,E\ln(E)\,.
\eea
Since $\chi(E)$ is always between $-\pi/2$ and $\pi/2$, all solutions 
to (\ref{econdlead}) for $l$ small have $k=0$. 
If $l$ is nonzero, then sufficiently
high in the spectrum the final term on the RHS of (\ref{econdlead})
becomes insignificant and the condition
reduces to $\chi(E)=-\pi l$. 
This is illustrated in figure \ref{trigfig2}:
given that $\epsilon$ is small, $\rho$ behaves as a slow
mode in $E$, and $\phi$ as a fast mode, and so 
the point $1+\rho\,e^{i\phi}$
moves around the circle of approximately-constant radius $\rho$, this
radius slowly decreasing as $E$ increases. 
Eigenvalues occur when this point crosses the ray from the origin
with argument $-\pi l$. Two such rays are shown on the figure: one, the
upper dashed line, is for $l<0$, and the other, the lower dashed line,
is for $l>0$.

\medskip

\[
\begin{array}{c}
\includegraphics[width=0.29\linewidth]%
{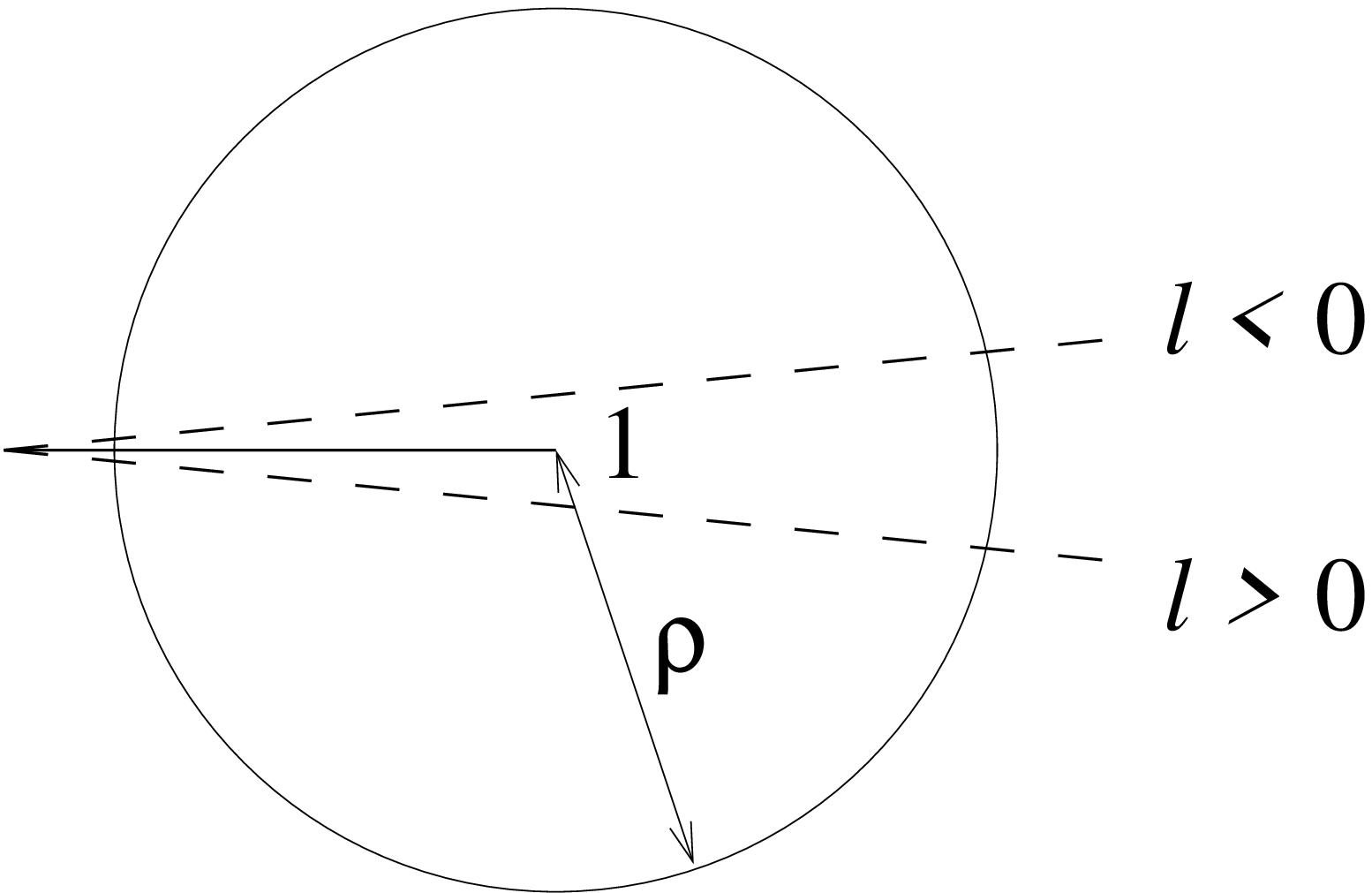}\\[9pt]
\parbox{0.6\linewidth}{\small
Figure \protect\ref{trigfig2}:
The approximate quantisation condition for $l\neq 0$.}
\end{array}
\refstepcounter{figure}\label{trigfig2}
\]

If $\rho$ becomes too small, the circle and the ray cease to
intersect and so the eigenvalue condition
can never be met, no matter what 
(real) value is taken by the fast mode $\phi$\,. This means
that the eigenvalues 
have merged to become complex. The critical value of $\rho$ at which
this first occurs is
$|\sin(\pi l)|$\,; hence, the merging of eigenvalues occurs
asymptotically along the curve
\eq
E=-\frac{8\ln|\sin(\pi l)|}{\pi^2|\epsilon|}~.
\label{lcurve}
\en
The curved dotted lines on 
figures~\ref{numerics}a,
\ref{numerics}b and
\ref{numerics}d 
allow (\ref{lcurve}) to be compared
with the behaviour of the exact levels.

We can also predict which eigenvalues pair off, by thinking about the
fast mode $\phi$. If $E$ is kept approximately fixed and $|\epsilon|$
is increased (since $\epsilon<0$, this corresponds to decreasing $M$),
then from (\ref{rhoform}) $\rho$ decreases while $\phi$ is unchanged.
This makes it easy to see that it is pairs of eigenvalues 
associated with neighbouring crossings of the ray $\arg(z)=-\pi
l$ on the {\em same} side of the real axis which merge. Now as $E$
increases from zero, the phase $\phi$ increases from $-\pi/2-\pi l$,
meaning that the point $1+\rho\,e^{i\phi}$ starts below the real axis.
If $l<0$, it first crosses into the upper half plane before 
encountering the ray $\arg(z)=-\pi l$, which it then does twice before 
crossing the real axis again. Hence the lowest level is paired with
the second level, the third with the fourth and so on. If on the other hand
$l>0$, then starting again from $E=0$ the point $1+\rho\,e^{i\phi}$
encounters the ray $\arg(z)=-\pi l$
{\em once}, below the real axis, then moves anticlockwise
through the upper half $z$-plane before the next two
crossings of the ray, both below
the real axis, occur. Hence the connectivity of the levels is
reversed, with the second and third, fourth and fifth, and so on,
levels pairing off. This is exactly as observed in our earlier
numerical work, now predicted analytically.

Strictly speaking, we have just found the connectivities that the
levels would have were they quantised by the nonperturbative term
alone, since we discarded the final term on the RHS of
(\ref{econdlead}) before commencing the analysis. 
However, if $E$ is taken large enough before
$\epsilon$ is sent to zero, then this term dominates and so our
argument also predicts the connectivity of the true level-mergings, high
up the spectrum. Lower down, the behaviour of the true levels may
differ from that of the nonperturbative-only approximation, as 
can be seen by
comparing figures \ref{pertcomp}b and \ref{pertcomp}c. This
corresponds to the recombination of levels, and simple
estimate of where it occurs can be obtained by finding the
intersection of the curve (\ref{lcurve}) with the curve
$\frac{\pi}{3}(l{+}\frac{3}{2})(l^2{-}\frac{1}{4})\,|\epsilon|E^{-2}=\pi 
l$,
which approximates the energy level found by quantizing according to
the perturbative term alone, shown for $l=-0.001$ in figure \ref{pertcomp}a.
For small $l<0$, this yields
\eq
E_{\rm recomb}(l)=\left(\frac{\ln|\pi l|}{\pi^2\,l}\right)^{1/3}.
\en
For $l=-0.001$\,, $E_{\rm recomb}\approx 8.36$, which matches well
the behaviour shown in figures \ref{numerics}c and \ref{pertcomp}.

It remains to treat the marginal case, $l=0$. This is much more
delicate. Easiest to discuss is the connectivity of the levels. 
The last term in (\ref{econdlead}) functions much as the term $-\pi l$
did before
from the point of view of the fast mode $\phi$; given its sign, this
means that the connectivity of levels for $l=0$ is the same as that
for $l>0$, which is indeed what was observed numerically.
(A similar argument explains why the connectivity is reversed below
$E=E_{\rm recomb}$ for $l<0$.)
To estimate where the level merging occurs, note that the critical
value of $\rho$ is now $\sin(\frac{\pi}{8}|\epsilon|E^{-2})$.
For large $E$ and small $\epsilon$ we must therefore solve
\eq
e^{-\frac{\pi^2}{8}|\epsilon|E}=\fract{\pi}{8}|\epsilon|E^{-2}
\en
to find the critical value of $E$. The solution can be expressed
in terms of the Lambert W function $W(x)$, which 
is defined to satisfy $W(x)e^{W(x)}=x$.  (More information on this
function can be found in \cite{lambertrefs};
it is related to the tree function $T(x)$ by
$W(x)=-T(-x)$.) Explicitly,
\eq
E=-\frac{16}{\pi^2|\epsilon|}\,W_{-1}%
\Big(-\frac{\pi^{5/2}|\epsilon|^{3/2}}{32\sqrt{2}}\Big)
\label{lcurve2}
\en
where the subscript $-1$ selects the relevant branch of the W
function (see \cite{lambertrefs}). Now using the asymptotic
$T_{-1}(x)=-W_{-1}(-x)\sim \ln(x^{-1}) + \ln(\ln(x^{-1}))$ for $x\to 0^+$\,,
the curve
(\ref{lcurve}), which diverges as $l\to 0$, should be replaced by
\eq
E\sim
\frac{16}{\pi^2|\epsilon|}\left[
\ln\Big(\frac{32\sqrt{2}}{\pi^{5/2}|\epsilon|^{3/2}}\Big)
+\ln\ln\Big(\frac{32\sqrt{2}}{\pi^{5/2}|\epsilon|^{3/2}}\Big)
\right].
\label{newasymp}
\en
This corrects the result of 
\cite{Bender:1998gh}, equation
(\ref{trancas}) above, and
shows that while the two-by-two truncation used in that paper
gives a
qualitative understanding of the situation, is too crude to capture precise
asymptotics.
However, even with the second logarithm included in (\ref{newasymp}), 
or using the W function directly, 
the approach of the level-mergings to their asymptotic locations for
$l= 0$ is much slower than that found for nonzero values of $l$.
The dotted line on figure~\ref{numerics}c
compares (\ref{lcurve2})
with the exact levels, obtained by solving the nonlinear integral
equation.

\resection{Asymptotics for the radial problem}
\label{Qasymp}
The approach of the previous sections can also be applied
to the more-standard radial problem
(\ref{hprob}).
The analysis is very similar to that already given,
though simpler because the subtleties of the second determination do
not apply.

The large-$\theta$ expansion of the first-determination kernel
(\ref{kerdef}) is
\begin{equation}
\varphi(\theta)\sim -
\sum_{n=1}^{\infty}\frac{1}{\pi}\cot(\fract{\pi}{2M}(2n{-}1))%
\,e^{-(2n-1)\theta} +
\sum_{n=1}^{\infty}\frac{M}{\pi}\tan(\pi Mn)%
\,e^{-2Mn\theta}
\label{kerdefexp}
\end{equation}
and so for large real $\theta$\,,
\begin{eqnarray}
f(E)&\sim& i\pi(l{+}\fract{1}{2})-
i\sum_{n=0}^{\infty}\frac{b_n}{\pi}\cot(\fract{\pi}{2M}(2n{-}1))%
E^{-\frac{(M{+}1)}{2M}(2n-1)} \nn\\[4pt]
&&\qquad\qquad\qquad {~}+
i\sum_{n=1}^{\infty}\frac{Mc_n}{\pi}\tan(\pi Mn)%
E^{-(M{+}1)n}\,,
\label{fexp}
\end{eqnarray}
where $b_n$, $c_n$ and $m$ are as before, and we defined
$b_0=-\pi\tan(\frac{\pi}{2M})m$ so as to absorb the
exponentially-growing part of the asymptotic into the first series. 
The condition for $E$ to be an eigenvalue of the radial problem can be
written as
\eq
f(E)=-(2k{+}1)\pi\,i\,,\quad k\in \Z
\en
Written directly in terms of
the conserved charges, and setting $I_{-1}:=1$,
the expansion coefficients of $f$ are
\begin{eqnarray}
{}-i\,\frac{b_n}{\pi}\cot(\fract{\pi}{2M}(2n{-}1))
&=& 
i\,(-1)^n\frac{\sqrt{\pi}\,\Gamma\Big(1{-}\frac{(2n-1)}{2M}\Big)}%
{\Gamma\Big(\frac{3}{2}{-}n{-}\frac{(2n-1)}{2M}\Big)}%
\frac{(4M{+}4)^n}{(2n{-}1)\,n!}\,I_{2n-1}\,; 
\label{flocalexpcoeff}\\[4pt]
i\,\frac{Mc_n}{\pi}\tan(\pi Mn)
&=&-2i\,
\sin(\pi Mn)
\frac{
2^{2Mn}
}{(M{+}1)^{2n}}\,\widetilde H_n\,.
\label{fnonlocalexpcoeff}
\end{eqnarray}

Notice that (\ref{fexp}) is a double series, unlike the
single series 
(\ref{attrexp}) and
(\ref{repexp})
found for the lateral problems,
and that its form does not change as we move
between the attractive ($M>1$) and the repulsive ($M<1$) regimes.
For $M<2$ the steepest-descent contour lies inside the first
determination, at least for large $E$, and so, unlike for the
PT-symmetric problem discussed above, there is no need to add a
non-perturbative term.
The double series could also have been recovered directly from
the asymptotics of $Q(\theta)$ found in \cite{Bazhanov:1996dr},
given the ODE/IM correspondence. Taking suitable care about the
branch choices, the same applies to the series parts of the expansions
for the lateral problems
obtained in the last section, but the beyond all
orders contributions would have been missed.

For the radial problem, the WKB method should
work for all values of $M$, and
it is natural to try to recover the asymptotics from a direct
analysis of the differential equation. In the rest of this
section we show how the
contributions to (\ref{fexp}) from local conserved quantities can be
found using an all-orders WKB approximation, and give some hints as to
the origins of the nonlocal parts.
Even for the local part of the discussion, there
are some interesting features,
as the application of the WKB technique to radial problems is
tricky~\cite{Langer,Feldman:1978si,Seetharaman}.

A WKB treatment of the radial problem might
begin by writing (\ref{hprob}) as
\eq
\left[-\frac{d^2}{dx^2}+\overline Q(x)\right]y=0~~,\quad
\overline Q(x)= x^{2M}-E+ \frac{l(l+1)}{x^2} \,.
\label{pot}
\en
For $l$ and $E$ positive, $\overline Q(x)$ has
two simple zeroes on the positive $x$
axis: one, $x_0$, near $x=0$ and the other, $x_1$, near $x=E^{1/2M}$.
The leading WKB quantisation condition 
would then be
$2\int_{x_0}^{x_1}\sqrt{-\overline Q(x)}\,dx=(2k{+}1)\pi$\,, 
$k\in\Z$\,.
However, this does not give
good results. The reason, first stressed by Langer \cite{Langer}, is
that the
WKB approximation derived from (\ref{pot}) breaks down near $x=0$.
Instead, Langer suggested making a preliminary change of variable
\eq
x= e^{\la}
\label{var}
\en
and gauge transformation
\eq
y(x)=e^{\la/2} \psi(\la)
\label{gauge}
\en
so that (\ref{pot}) becomes
\eq
\left[-\frac{d^2}{dz^2}+\widetilde Q(z)\right] \psi=0~~,\quad
\widetilde Q(\la)=e^{(2M+2) \la }- E e^{2 \la }  + (l{+}1/2)^2\,.
\label{leqn}
\en
The quantisation condition for the modified problem is
\eq
2\int_{z_0}^{z_1} \!\sqrt{-\widetilde Q(\la)}
\,d \la =(2k{+}1) \pi \,,~~~~k\in\Z\,,
\label{modquant}
\en
which, changing the variable back to $\ln z=x$,
amounts to the trading of $l(l{+}1)$ with $\lambda^2=(l+1/2)^2$
in the original potential. This is sometimes called
the Langer modification, and it allows the energy levels of the
radial harmonic oscillator and the radial Coulomb potential
to be recovered exactly. However, away from these points, the
generalisation to incorporate higher-order WKB corrections
is difficult.  In particular, the leading
$l$-dependence in (\ref{fexp}), namely the term
$i\pi(l{+}\frac{1}{2})$, only emerges once an all-orders resummation
has been performed~\cite{Seetharaman}.

The problem is that $l$ and $E$ are
badly tangled up in formulae such as (\ref{modquant}). To get round
this, we take the Langer-transformed
equation (\ref{leqn}) and change variable $\la \rightarrow \gamma
\la$.
Transforming back to the variable $x$, using
(\ref{var}) and (\ref{gauge}) in reverse,
 we find
\eq
\left[ - {d^2 \over d x^2}  + \gamma^2 x^{(2M+2) \gamma-2}- E \gamma^2
x^{2 \gamma -2 } + {\tilde{l} (\tilde{l} +1) \over x^2} \right] y =0
\label{tr}
\en
where
\eq
\tilde{l}= \gamma \lambda -1/2~.
\en
If $\gamma=1/(2 \lambda)$ then $\tilde l=0$,
and, changing variable to $t=E^{-\lambda/M}x$,
(\ref{tr}) simplifies to
\eq
\left[ - \varepsilon^2{d^2 \over d t^2}  + Q(t) \right] y(t) =0
\label{simp1}
\en
where
\eq
 {Q}(t)= \frac{1}{4\lambda^2}\, t^{1/\lambda-2}
(t^{M/\lambda}-1)\,,\quad \varepsilon=
{E^{-(M{+}1)/2M}} \,.
\label{newpot}
\en
A key feature of this equation is that
the $E$-dependence, contained in $\varepsilon$, has been
factored out of the
transformed potential $Q(t)$. In fact,
$\varepsilon$ takes the role of the variable which
organises the whole higher-order WKB series, in eq.~(\ref{wkbsol})
below. This means that a systematic WKB treatment of
the transformed equation should yield the asymptotic behaviour of the
energy levels as $E\to\infty$ at fixed $M$ and $l$, which is just what we
want.
The contrast with (\ref{leqn}) accounts for
the difficulties encountered when trying to work directly with the
Langer-transformed equation. The simplification has a price,
though: for $l\neq 0$ the turning point which had been near the
origin is replaced by a singularity of order $1/\lambda-2$ exactly
at $x=0$.
This requires special treatment, which we illustrate by calculating
the leading correction explicitly by asymptotic matching.

To build a solution which to leading (`physical optics')
asymptotic approximation solves the transformed
eigenvalue problem, the $t$ axis can be split into four regions,
defined for $\varepsilon\to 0^+$ as follows\footnote{As in
\cite{bender:book}, $a\ll b ~(\varepsilon\to 0^+)
\leftrightarrow a/b\to 0$ as $\varepsilon\to 0^+$}:
\bea
\mbox{Region \,I\,~} &:& t>1,~ 
|t{-}1|\gg\varepsilon^{2/3}\nn\\
\mbox{Region II~} &:& |t{-}1|\ll 1 \nn\\
\mbox{Region III} &:& 1>t>0,~ |t|\gg\varepsilon^{2\lambda},~
|1{-}t|\gg\varepsilon^{2/3}\nn\\
\mbox{Region IV} &:& t>0,~|t|\ll 1
\eea
In regions I and III, the physical optics approximation is good, while
regions II and IV need separate treatment.
In region I, the decaying WKB solution is
\eq
y\php_{\rm I}(t) = \frac{1}{[Q(t)]^{1/4}}
\exp \left[ - \frac{1}{\varepsilon}\int_1^{t} \sqrt{Q(u)}\,du \right].
\en
This can be continued through region II, which contains the simple
turning point at $t=1$,
using an Airy function in the standard way~\cite{bender:book}, to find
$y(t)\sim y\php_{\rm III}(t)$
in region III, where
\eq
y\php_{\rm III}(t) = \frac{2}{[-Q(t)]^{1/4}}
\cos \left[ - \frac{1}{\varepsilon}\int_t^{1} \sqrt{-Q(u)}\,du +
\frac{\pi}{4} \right].
\en
Taking $t$ near zero, but still in 
region III, this behaves as
\bea
y\php_{\rm III}(t) &=&
\frac{2}{[-Q(t)]^{1/4}}
\cos \left[ - \frac{1}{\varepsilon}\int_0^{1} \sqrt{-Q(u)}\,du +
 \frac{1}{\varepsilon}\int_0^{t} \sqrt{-Q(u)}\,du +
\frac{\pi}{4} \right]\nn\\[3pt]
&\sim&
2\sqrt{2\lambda}\,t^{\frac{1}{2}-\frac{1}{4\lambda}}\,
\cos \left[ -\frac{1}{\varepsilon} \int_0^{1} \sqrt{-Q(u)} \,du +
 \frac{1}{\varepsilon}\, t^{\frac{1}{2\lambda}}+
\frac{\pi}{4} \right]\,.
\label{asone}
\eea
This solution
breaks down as $t\to 0$, and must be matched to
a solution of the approximate ODE in region IV, which is
\eq
\left[ - {d^2 \over d t^2}  - 
\frac{1}{4\lambda^2\varepsilon^2}\,t^{\frac{1}{\lambda}-2} \right] y(t)
=0
\en
Imposing $y\to 0$ as $t\to 0$, the appropriate solution is
\eq
y\php_{\rm IV}(t)=\beta
\sqrt{t}\,J_{\lambda}\big(\fract{1}{\varepsilon}\,%
t^{\frac{1}{2\lambda}}\big)\,,
\label{yivdef}
\en
where $J_{\lambda}$ is a Bessel function, and
$\beta$ a so-far arbitrary normalisation constant.
(The behaviour as
$\frac{1}{\varepsilon}\,t^{\frac{1}{2\lambda}}\to 0$ is
$y\php_{\rm IV}(t)\sim 
\beta t\,(2\varepsilon)^{-\lambda}/\Gamma(\lambda{+}1)$,
as follows from the $\xi\to 0$ asymptotic
$J_{\lambda}(\xi) \sim (\xi/2)^{\lambda} / \Gamma(\lambda{+}1)$.)
The large-$\xi$ formula
$J_{\lambda}(\xi) \sim \sqrt{ 2 \over \pi \xi} \cos
\left [ \xi-{\lambda \over 2} \pi  - {\pi \over 4} \right ]$
then implies
\eq
y\php_{\rm IV}(t)\sim
\beta
\sqrt{\frac{2\varepsilon}{\pi}}\,t^{\frac{1}{2}-\frac{1}{4\lambda}}\,
\cos \left(
\frac{1}{\varepsilon}\, t^{\frac{1}{2\lambda}} - \frac{\lambda \pi}{2}-
\frac{\pi}{4} \right)
\quad\mbox{for}~~
t\gg \varepsilon^{2\lambda}\,.
\label{astwo}
\en
For $E$ to be an eigenvalue, this should match the small-$t$
form of $y\php_{\rm III}(t)$ given by (\ref{asone}).
This requires
\eq
\frac{1}{\varepsilon}\int_0^1\sqrt{-{Q}(u)}\,du=k\pi+
\frac{\lambda\pi}{2}+
\frac{\pi}{2}~,\quad 
\beta= 2\sqrt{\frac{\pi\lambda}{\varepsilon}}\, (-1)^k ,\quad
k\in\Z\,.
\en
Restoring the original variables and substituting
$w=u^{\frac{M}{\lambda}}$ results in a beta-function integral and
the quantisation condition
\eq
\frac{1}{M}
B(\fract{1}{2M},\fract{3}{2})
\,E^{\frac{M{+}1}{2M}}
=(2k+l+\fract{3}{2})\pi~,\quad  k\in\Z\,.
\en
Using (\ref{betaf}), the condition
implied by
(\ref{fexp}) is recovered to next-to-leading order\footnote{%
Following Langer \cite{Langer}, the same result can also be
obtained starting from the approximation
$y(t) \sim    [- Q(t)]^{-1/4} \xi^{1/2} \left(\alpha
J_{-\lambda}(\xi)+ \beta J_{\lambda}(\xi)\right)$ in the region
$t\approx 0$, with
$\xi= \int_0^{t} \sqrt{-Q(u)} du$\,, and the boundary condition
at $t=0$ requiring $\alpha=0$.}. Notice that the full
$l$-dependence has been recovered directly, without any need for
resummation.

Encouraged by this result, it is natural to see how much more of the
asymptotic expansion
found using the ODE/IM correspondence can be recovered by
incorporating higher-order WKB corrections.
Suppose that (\ref{simp1}) has a  solution of the form
\eq
y(t)=\exp \left[\frac{1}{\varepsilon}\sum_{n=0}^{\infty}
\varepsilon^nS_n(t) \right] .
\label{wkbsol}
\en
For (\ref{simp1}) to be satisfied order-by-order in $\varepsilon$,
the derivatives 
$S'_n(t)$ must obey the following recursion relations:
\eq
S'_0(t)= - \sqrt{Q(t)}~,~~~2 S'_0 S'_n +\sum_{j=1}^{n-1} S'_j S'_{n-j}
+S''_{n-1}=0~~(n \ge 1)\,.
\label{rec}
\en
The first few terms of the solution are
\bea
S'_0 &=& - \sqrt{Q}~, \nn \\[5pt]
S'_1 &=& - \frac{\,Q'}{4Q}~, \nn \\[2pt]
S'_2 &=& - \frac{1}{48} \left( \frac{Q''}{Q^{3/2}} +5
\frac{d}{dz} \left[\frac{Q'}{Q^{3/2}} \right]\right),\nn\\[3pt]
S'_3 &=& - \frac{~Q''}{16Q^2}+\frac{5(Q')^2}{64Q^3}\,=\,
\frac{d}{dz}\left[\frac{5(Q')^2}{64Q^3}-\frac{Q''}{16Q^2}\right],
\eea
Keeping just $S_0$ and $S_1$ constitutes the
physical optics approximation employed above;
further terms are very easily obtained 
using, for example, Mathematica.
Near to zeros of $Q(t)$ -- the so-called turning points -- the
approximation breaks down and further work is needed to find the 
connection formulae for the continuation of WKB-like solutions of
given order
from one region of non-vanishing $Q$ to another, just as was done
above.
Solutions found by continuation away from the two boundary conditions
must then be matched to find the condition for $Q$ to be such that
the differential equation has an acceptable solution.
This quickly becomes quite complicated, as
exemplified by the calculations in section 10.7 of~\cite{bender:book}.

In cases where $Q$ is entire with
just a pair of well-separated simple zeros on the real axis,
Dunham~\cite{Dunham} 
found a remarkably simple formulation of the final condition, 
valid to all orders:
\eq
\frac{1}{i}\oint \sum_{n=0}^{\infty} \varepsilon^{n-1}
S'_n(z)\, dz =2k \pi\,,~~~~k\in\Z
\label{quanta}
\en
where the contour encloses the two turning points; it closes
because for such a $Q$
all of the functions $S'_n$ derived from (\ref{rec}) 
are either entire, or else have a pair of square root branch
points which can be connected by a branch cut along the real axis.
Notice that the contour can be taken to lie far from the two turning
points, where the WKB series breaks down, and so there is no need to
worry about connection formulae.
All of the terms $S'_{2n+1}$, $n\ge 1$, turn out to be total
derivatives and can therefore be discarded,
while $\frac{1}{2i}S'_1=-\frac{1}{8i}Q'/Q$ and contributes a 
simple factor of $\pi/2$ when integrated round the two zeros of $Q$,
irrespective of any other details.
Dunham's condition is therefore
\eq
\frac{1}{i}\oint \sum_{n=0}^{\infty}\varepsilon^{2n-1}
 S'_{2n}(z)\, dz =(2k{+}1) \pi\,,~~~~k\in\Z\,.
\label{quant}
\en

However, this method is only directly relevant to the current
problem if the angular momentum is zero, and $M$ is an integer.
Extensive discussions of these cases can be found in \cite{Bender:dr},
and it is straightfowardly checked that the results found there match
the expansion we obtained above using
the ODE/IM correspondence.
Note that for $M\in\Z$ there are no `nonlocal' contributions
to the asymptotic (\ref{fexp}), so the WKB series gives the complete
answer.

For more general cases there is only one simple
turning point, the other being replaced by the singularity at $z=0$,
and the analysis just given does not apply.
Nevertheless, the $E$-dependence
of (\ref{fexp}) together with the match with the results of
\cite{Bender:dr} at special points suggests
that the contributions to the asymptotic related to the
{\em local\/} integrals of motion might still
be obtained from a WKB series of the form (\ref{quant}), suitably
treated.  The main difficulty is 
the fractional singularity at $z=0$\,, which
prevents the contour in Dunham's condition from closing.
However, away from regions about $z=0$ and $z=1$
which are vanishingly small as $\varepsilon\to 0$,
the all-orders WKB solution provides a good
approximation to the true wavefunction.
As an {\em ad hoc}\/
measure, we replace the closed Dunham contour
in (\ref{quanta}) 
by a contour ${\cal C}$ which starts just below the origin,
passes once round the turning point at $z=1$, and returns to a
point just above the origin. Since all of the terms $S'_{2n+1}$, $n\ge
1$, are total derivatives of functions which are single-valued around
$z=1$, they again make no contribution even though the contour is no
longer closed. Therefore we replace Dunham's condition (\ref{quant})
by
\eq
{1 \over i} \int_{\cal C} \sum_{n=0}^{\infty}\varepsilon^{2n-1}
 S'_{2n}(z)\, dz =(2k+\frac{1}{2}+2\delta)
\pi\,,~~~~k\in\Z
\label{rquant}
\en
where $S'_{2n}$ is obtained from (\ref{rec}) using the $Q$ given by
(\ref{newpot}), and
the factor $2\delta$ allows for a possible phase-shift caused by 
the singularity at the
origin. From the initial asymptotic matching calculation, the leading
(constant) part of $2\delta$ is $\lambda+1/2$, but 
further $E$-dependent corrections can be expected -- we shall return
to this issue below.

For concrete calculations, it is convenient to collapse
the contour ${\cal C}$ 
onto real axis, and the square root singularities of the integrands at
$z=1$ then allow each integral to be replaced by
$2 \int_0^1  S'_{2n}(t)\,dt$\,. A difficulty with this
procedure is that the divergences in the WKB series at $z=1$ are no
longer avoided. We remedied this by 
multiplying each integrand
$S'_{2n}(t)$ by $(t^{M/\lambda}-1)^{\kappa}$ to force convergence, 
doing the definite
integrals -- still possible in closed form -- and then setting $\kappa=0$ 
at the end. It is a simple matter 
to mechanise the calculation with a few
lines of Mathematica code, and,  using (\ref{flocalexpcoeff}) in
reverse, 
the values of the local charges
$I_1$, $I_3$ and $I_5$ given in (\ref{loccharges}) are easily reobtained;
we also reproduced the formulae for $I_7$ and $I_9$ quoted in
\cite{Bazhanov:1994ft,Bazhanov:1996aq}, though these are rather too
lengthy to be worth repeating here.
It is interesting that the values of the conserved charges in the
quantum field theory can be recovered by such
relatively-straightforward manipulations of the ordinary differential
equation. 

Finally, the phase shift $\delta$ should be analysed.  We 
claim that this
is related to the part of the asymptotic due to non-local
conserved charges. To see why, focus on the
behaviour near the origin by changing variables one
more time in (\ref{simp1}), to $u:=\varepsilon^{-2\lambda}t$. The
equation becomes
\eq
\left[ - {d^2 \over d u^2}  -
\frac{1}{4\lambda^2}\,u^{1/\lambda-2}+
\frac{1}{4\lambda^2}\,\varepsilon^{2M}\,u^{(M{+}1)/\lambda-2}\right]
y(u)=0\,.
\label{simp2}
\en
Treating the final term as a perturbation
results in a series of corrections to $y(u)$ as powers of 
$\varepsilon^{2M}=E^{-(M+1)}$. Comparing
with (\ref{fexp}), this is exactly the structure of the second series
of terms emerging from the ODE/IM correspondence, encoding 
the values
of the non-local conserved charges in the integrable model.

In more detail, the perturbative treatment begins with
the unperturbed equation
\eq
\left[ - {d^2 \over d u^2}  -
\frac{1}{4\lambda^2}\,u^{1/\lambda-2}
\right]
y(u)=0\,.
\en
Two independent solutions $j$ and $n$ are
\eq
j(u)=\sqrt{u}\,J_{\lambda}(u^{\frac{1}{2\lambda}})~~,\quad
n(u)=\sqrt{u}\,Y_{\lambda}(u^{\frac{1}{2\lambda}})\,.
\en
Their Wronskian is $W[j,n]:=j(u)n'(u)-j'(u)n(u)=\frac{1}{\pi\lambda}$
and using this fact it can be checked (as in for example
\cite{Landau,Calogero})
that, with the boundary condition $y(u)\sim j(u)$ as $u\to 0$, 
the full differential equation (\ref{simp2})
is equivalent to the integral equation
\eq
y(u)=j(u)+\varepsilon^{2M}\!\int_0^u G(u|s)\,v(s)y(s)\,ds
\label{born}
\en
where
\eq
G(u|s)=\pi\lambda\left[n(u)j(s)-n(s)j(u)\right]
\en
and
\eq
v(u)=\frac{1}{4\lambda^2}\,u^{(M{+}1)/\lambda-2}\,.
\label{pertpot}
\en
Formally, (\ref{born}) can be solved by iteration. The zeroth order
result is proportional to $y\php_{\rm IV}(t)$, defined in
(\ref{yivdef}); later terms give the Born series for the problem. 

To extract the corrected phase shift, the RHS of
(\ref{born}) can be rewritten as
\eq
y(u)=j(u)\left[1-
\varepsilon^{2M}\!\int_0^u \!\pi\lambda\,n(s)\,v(s)y(s)\,ds\right]
+n(u)\,
\varepsilon^{2M}\!\int_0^u \!\pi\lambda\,j(s)\,v(s)y(s)\,ds~.
\label{borntwo}
\en
The large-$u$ asymptotics
\eq
j(u)\sim \sqrt{\fract{2}{\pi}}\,u^{\fract{1}{2}{-}\fract{1}{4\lambda}}\,
\cos\big(u^{\fract{1}{2\lambda}}{-}\fract{\lambda\pi}{2}{-}
\fract{\pi}{4}\big)~,~~
n(u)\sim \sqrt{\fract{2}{\pi}}\,u^{\fract{1}{2}-\fract{1}{4\lambda}}\,
\sin\big(u^{\fract{1}{2\lambda}}{-}\fract{\lambda\pi}{2}{-}\fract{\pi}{4}\big)
\en
then yield
\eq
y(u)\sim
\sqrt{\fract{2}{\pi}}\,u^{\fract{1}{2}-\fract{1}{4\lambda}}\,
A(u)
\cos\big(u^{\fract{1}{2\lambda}}{-}\pi\delta\big)
\label{phasympt}
\en
where
\eq
\tan\left(\pi\delta{-}\fract{\lambda\pi}{2}{-}\fract{\pi}{4}\right)=
\varepsilon^{2M}\!\int_0^u \!\pi\lambda\,j(s)\,v(s)y(s)\,ds
\Bigg/
\left[1-
\varepsilon^{2M}\!\int_0^u \!\pi\lambda\,n(s)\,v(s)y(s)\,ds\right].
\label{phase}
\en
In problems with localised perturbing potentials, the integrals in 
(\ref{phase}), and also the function $A(u)$ in 
(\ref{phasympt}), tend to finite limits as $u\to\infty$.
Replacing $y(s)$ on the RHS of (\ref{phase}) by the Born series
generated by (\ref{born}) then gives the perturbative expansion of the
phase shift.
This must be treated with some caution here, since the perturbing potential
(\ref{pertpot}) in general grows at infinity. However, at least at
first order -- the so-called Born approximation -- 
it is possible to extract sensible results.

The leading behaviour of (\ref{born}) is simply $y(u)\approx j(u)$. At 
this order there is no contribution from the denominator in
(\ref{phase}), and so the once-corrected phase shift is, formally, 
\bea
\delta 
&=& \frac{\lambda}{2}+\frac{1}{4}+
\varepsilon^{2M}\int_0^{\infty}\lambda\,v(s)j(s)^2\,ds+\dots\nn\\[3pt]
&=& \frac{\lambda}{2}+\frac{1}{4}+
\frac{1}{2}
E^{-(M{+}1)}\!
\int_0^{\infty}
x^{2M{+}1}
J_{\lambda}(x)^2\,dx+\dots
\eea
The last integral converges for $M+1+\lambda>0$, $2M+1<0$\,, and in this
region its value (found using Mathematica, or, for example, page 237 of
\cite{SF}) is
\eq
\int_0^{\infty} x^{2M{+}1} J_{\lambda}(x)^2\,dx=
\frac{\Gamma(-M{-}\frac{1}{2})\Gamma(M{+}1{+}\lambda)}%
{\sqrt{2\pi}\,\Gamma(-M)\Gamma(-M{+}\lambda)}~.
\en
Using (\ref{fnonlocalexpcoeff}), 
this reproduces the value of $\widetilde H_1$ given 
by (\ref{nlval}) above.

The agreement with previous results provides some retrospective
justification for our procedure, though we cannot rule out the appearance
of subtleties at higher orders. Clearly, a more rigorous and
systematic treatment would be desirable, especially given the
difficulties in evaluating the eigenvalues of the nonlocal charges
directly within the quantum field theory. However, we shall leave further 
investigation of this point for future work.

\resection{Conclusions}
\label{concl}
We have given a concrete application of the ODE/IM
correspondence by showing how it can be used to obtain an analytic
understanding of the level mergings in 
the model introduced by Bender and Boettcher, one of the
longest-studied examples of $\PT$ symmetric quantum mechanics. 
The subtle mixture of perturbative and nonperturbative effects
contributing to the recombination of levels in the generalisation of
the model to include a centrifugal term is particularly
striking, and shows once again the surprising richness of $\PT$ symmetry
as a source of interesting problems in mathematical physics.

There are many questions left unanswered by our work, and we finish by
mentioning just a few potentially interesting directions for further
investigations.

First of all, the origin of the perturbative and nonperturbative terms
for $M<1$ should be understood through more standard differential
equation techniques. For the nonperturbative term responsible for the
level-mergings, we expect that the complex WKB method will play a
role, as in \cite{BBMSS}.  
This could enable the study of models not 
treatable with the (less conventional)
methods described in this paper. 
An alternative strategy for systems with potentials of the form
$P(x)/x^2$ with $P(x)$ a polynomial would be to use 
the set of functional relations (the fusion hierarchy) satisfied by
the spectral determinant $T(E)$, and this could also be explored.  

Our discussion of the nonperturbative term was restricted to
$M\approx 1$; as mentioned above, when $M$
goes beyond $1/2$ or $2$, the pole responsible for this term 
no longer crosses the steepest-descent contour and the asymptotic
changes. This effect turns out to be most marked near to $M=1/2$ for
$l=0$, and in this region some further smoothing is needed before
a good approximation for the energy levels can be obtained from the
integral equation. Relevant techniques for analogous problems have
been developed in \cite{berryhowls,boyd}, but we have yet to apply
them to the current situation. It would be interesting to see whether
the delicate asymptotic for the diverging ground state energy
calculated in \cite{Bender:1998ke,Bender:1998gh}
could be recovered by such methods.

The perturbative parts of the expansions for both the radial and the
lateral problems are especially interesting for the ODE/IM
correspondence, since they encode the values of conserved
charges in the integrable quantum field theories.  
Our discussion of the relationship between the Born series and the
nonlocal conserved charges at the end of section 
\ref{Qasymp} was rather preliminary, and 
it would be nice to make the analysis fully rigorous
and to push it to higher orders. In this respect, the
sophisticated calculations of spectral zeta functions of 
Chudnovsky, Chudnovsky and Voros \cite{Voros} are likely to be
relevant. The relationship between the local conserved charges and
the WKB series is much clearer, but still needs to be understood on a
more profound level. On the ODE side, efficient techniques have been 
developed for the evaluation of higher-order WKB contributions 
(see for example \cite{robnik}), and there now seems to be
scope to apply these methods to the study of integrable models.

Finally, it would be worthwhile to extend these considerations to other
integrable models, and other ordinary differential equations.
Some results in this direction, for local conserved charges,
can be found in \cite{Lukyanov:2003nj,Vitchev:2004iz}; two further
examples to study would be
the higher-order equations related to $SU(n)$ Bethe ansatz systems
discussed in \cite{Dorey:2000ma},
and the Schr\"odinger equations for excited states found in
\cite{Bazhanov:2003ni}.

\section*{Acknowledgments}
We would like to thank Carl Bender, Clare Dunning,
Chris Howls, Frieder Kleefeld,
Junji Suzuki, Andr\'e Voros and Miloslav Znojil
for useful conversations
and correspondence. RT  thanks the EPSRC for an Advanced Fellowship.
This work was partly supported by the EC network ``EUCLID", 
contract number HPRN-CT-2002-00325, and partly by a NATO grant
PST.CLG.980424. PED was also supported in part by a
JSPS/Royal Society grant and by the Daiwa Foundation, and thanks SPhT
Saclay for hospitality while this project was in progress.
%
%

\end{document}